# A Systematic Framework for Dynamically Optimizing Multi-User Wireless Video Transmission

Fangwen Fu, Mihaela van der Schaar
Electrical Engineering Department, UCLA
{fwfu, mihaela}@ee.ucla.edu

***Abstract*****—** In this paper, we formulate the collaborative multi-user wireless video transmission problem as a multi-user Markov decision process (MUMDP) by explicitly modeling the users' heterogeneous video traffic characteristics and time-varying network conditions. Our formulation further considers the existing coupling between the wireless users when determining the optimal cross-layer transmission strategies that should be adopted by the users to maximize their long-term system utility (i.e. video quality). These environment dynamics are often ignored in existing multi-user video transmission solutions. To comply with the decentralized architecture of multi-user communication scenarios, we propose to decompose the MUMDP into local MDPs, which can be autonomously solved by individual users, using Lagrangian relaxation. Unlike in conventional multi-user video transmission solutions stemming from the network utility maximization framework, the proposed decomposition enables each wireless user to individually solve its own *dynamic* cross-layer optimization and the network coordinator to update the Lagrangian multipliers (i.e. resource prices) based on not only current, but also future resource needs of all users, such that the long-term video quality of all users is maximized. However, solving the MUMDP requires statistical knowledge of the experienced environment dynamics, which is often unavailable before transmission time. To overcome this obstacle, we propose a novel online learning algorithm, which allows the wireless users to update their policies in multiple states during one time slot. This is different from conventional learning solutions, which update one state per time slot. The proposed learning algorithm can significantly improve the learning performance, thereby dramatically improving the video quality experienced by the wireless users over time. Our simulation results demonstrate the efficiency of the proposed MUMDP framework as compared to conventional multi-user video transmission solutions.

## I. INTRODUCTION

Due to their flexible and low cost infrastructure, wireless networks are poised to enable a variety of delay-sensitive multimedia transmission applications, such as videoconferencing, surveillance, telemedicine, remote teaching and training, and distributed gaming. However, existing wireless networks provide dynamically varying resources with only limited support for the Quality of Service (QoS) required by *delay-sensitive, bandwidth-intense and loss-tolerant* multimedia applications. Moreover, as multimedia applications continue to proliferate, wireless network infrastructures will often need to support multiple simultaneously running applications. Key challenges associated with the robust and efficient multi-user video transmission over wireless networks are the *dynamic* allocation of the scarce network resources among heterogeneous users experiencing different time-varying network conditions and traffic characteristics, and the *dynamic* adaptation at the individual users given their allocated network resources.

Existing wireless video transmission solutions can be broadly divided into two categories: single-user video transmission solutions, focusing on packet scheduling, error protection or cross-layer adaptation in order to maximize the receiving user's video quality [1]-[6], and multi-user video transmission, emphasizing multi-user resource allocation among multiple users simultaneously transmitting video and sharing the same wireless resources [14]-[18]. However, most existing solutions in both categories do not explicitly consider both the heterogeneous characteristics of the video traffic and the time-varying network conditions (e.g. time-varying channel conditions, dynamic multi-user channel access, etc.), thereby often leading to suboptimal performance for wireless media systems. For example, in the single-user video transmission category, most solutions employ Unequal-Error-Protection (UEP) techniques [7][8] to differentially protect the video packets based on their distortion impact and delay deadline. The work in [2] further proposes a rate-distortion optimized packet scheduling solution, which explicitly considers the video packets' dependencies by using a directed-acyclic-graph (DAG). This rate-distortion optimization method was subsequently improved in [4] by reducing the policy search space and in [6] by reducing the start-up latency for real-time video streaming applications. However, these solutions assume only simplistic underlying network (channel) models and they do not consider the adaptation of transmission parameters at the other layers of the network stack, besides the application layer. To deal with the dynamics in the wireless network, cross-layer adaptation methods [1][3][5] have been proposed to optimize on-the-fly the transmission parameters at various layers, based on current observations of the channel conditions. However, these cross-layer solutions are myopic and result in suboptimal performance because they do not account for the future channel conditions and video traffic.

In the multi-user video transmission category, many current techniques [14]~[18] are based on the network utility maximization (NUM) framework [13]. In the NUM framework, the basic assumption is that each user has a static utility function of the (average) allocated transmission rate (or QoS). For example, the authors in [15] simply consider the utility to be a function of the average allocated rate. The solutions in [14][16][17] (including our previous work in [18]) defined the utility function (i.e. the average video quality or distortion) as a function of the average rate and packet loss. To deal with the dynamic wireless channel conditions, the resource allocation among the multiple users is repeatedly performed to maximize the current video quality. However, these solutions only myopically maximize the video quality for all the users at the current time and do not predict the impact of the current resource allocation on the future video quality of all the users. This myopic solution leads to fluctuations in the long-term video quality. Therefore, it is crucial to judiciously allocate the limited resources to individual wireless video users such that their long-term utility (i.e. video quality) is maximized.

To address the abovementioned challenges associated with efficient multi-user video transmission over the time-varying wireless network, we propose a systematic framework [1] for dynamically and foresightedly optimizing the long-term video quality of multiple users coexisting in the same wireless channel. Under the proposed framework, unlike the existing single-user video transmission solutions, we explicitly consider when performing the cross-layer optimization for each wireless user, both the heterogeneous video traffic characteristics and the experienced time-varying network conditions. First, to characterize the heterogeneous video data, we define a traffic state for each user which considers the amount of data units (e.g. video frames or video packets) to be transmitted, their distortion impacts, and the dependencies between them at each transmission time. The traffic and network state (e.g. channel conditions) transitions characterize the environment dynamics experienced by each user. These dynamics are affected by the packet scheduling deployed by all the users and their resource acquisition. We then formulate the packet scheduling and resource acquisition for each user as an MDP. The deployed solution to the single-user video transmission problem using MDP can be found in our prior work [35].

[1] In order to facilitate the dynamic optimization of video transmission over time-varying wireless networks, we consider a time-slotted wireless transmission system in which the decision is made every time slot. The length of one time slot is determined based on how fast the network and video traffic are changed [18].

We further formulate the optimization of the packet scheduling and resource allocation over the dynamic multi-user video transmission system as an MUMDP [19] problem. Similar to the MDP formulation for single-user video transmission, the MUMDP formulation allows each user to make foresighted transmission decisions by taking into account the future impact of its current decisions on the long-term utilities of all the users.

Although the MUMDP problem can be solved in a centralized manner using conventional value or policy iteration algorithms or linear programming [19], this requires the network coordinator to know the dynamics experienced by each user and to solve a highly complex centralized MUMDP, involving a very large state space. Therefore, this solution leads to very high computation complexity and unacceptable communication overheads and incurred delays. Fortunately, the MUMDP is weakly-coupled [20], since the state transition of each user is coupled with that of other users only through the multi-user resource allocation at each time slot. We propose to decompose this weakly-coupled MUMDP problem using Lagrangian relaxation into multiple local MDPs each of which can be separately solved by the individual users. This decomposition is different from the conventional dual solutions [15] to the multi-user NUM-based video transmission problem in two ways: (i) instead of maximizing the static utility at each transmission time, our approach allows each wireless user to solve the dynamic cross-layer optimization problem (formulated as the local MDP), which is vital for the delay-sensitive video applications; (ii) instead of updating the Lagrangian multipliers only based on the current resource requirements of all users, our approach updates the multipliers based on not only the current but also future resource needs, such that the long-term video quality of all the users is maximized.

We would like to note that the Lagrangian relaxation using a scalar multiplier has been proposed in [20][21] to decompose a weakly-coupled MUMDP problem. However, the duality between the proposed relaxed problem and the original MUMDP is not analyzed in these existing works. In this paper, we mathematically derive the duality and propose a systematic way to compute the subgradient for updating the Lagrangian multiplier. Furthermore, based on our knowledge, we believe that this is the first attempt to formalize the multi-user video communication problem using MUMDP and decompose the MUMDP for autonomous, but collaborative, video users.

To solve the local MDP, each wireless user is required to know the transition probabilities of traffic states and channel states beforehand, which is often difficult to accurately characterize before transmission time, especially for video applications operating in dynamic multi-user networks. Hence, the local MDP cannot be solved in practice using the method in [35] and the Lagrangian multipler cannot be updated based on the subgradient method. Importantly though, the MUMDP framework provides the necessary foundations and principles for how the users can autonomously learn on-line to cooperatively optimize the global video utility when the dynamics are unknown. In practice, to deal with the unknown dynamics, each wireless user will deploy online reinforcement actor-critic learning [27], and the network coordinator will update the resource price dynamically using stochastic subgradient methods [26]. This approach has two advantages: (i) it does not require each user to know the statistical distribution of channel conditions and incoming video traffic; and, (2) the wireless user and the network coordinator need to perform only very simple computations during each time slot. Unlike conventional online learning algorithms [27], which often update the learning policy for only one state during each time slot, our proposed learning algorithm can update multiple states simultaneously during each time slot. As demonstrated by our simulation results, the proposed reinforcement learning solution significantly increases the learning performance, which leads to dramatic improvements in the received video quality.

The paper is organized as follows. Section II defines the traffic states, the state transition and the utility function for each wireless user at each time slot. Section III discusses the dynamic optimization for the single user. This is a local problem solved by the individual users in the multi-user video transmission problem. Section IV formulates the multi-user video transmission problem as an MUMDP. Section V presents how the MUMDP can be decomposed into multiple local MDPs using the Lagrangian relaxation method and develop the corresponding subgradient method to update the resource price. Subsequently, Section VI describes the proposed distributed online learning algorithm to deal with the unknown video characteristics and channel conditions. Section VII presents numerical results to validate the proposed framework. The conclusions are drawn in Section VIII.

## II. MODELS FOR HETEROGENEOUS VIDEO TRAFFIC

Unlike traditional traffic models [33], which only characterize the rate changes of video traffic, in this section we aim to develop a general model for representing the encoded video traffic with heterogeneous characteristics (e.g. various delay deadlines, distortion impacts, dependencies, etc.). Using this video traffic model, we will be able to dynamically optimize the resource acquisition and packet scheduling for video transmission over time-varying networks.

### A. Attributes of DUs

In this subsection, we discuss how the heterogeneous attributes of the video traffic[2] can be modelled. This will be used to define the traffic states at each time slot in Subsection B. The video data is encoded periodically using a Group of Pictures (GOP) structure as in [22][23], which lasts a period of $T$ time slots. The video frames within one GOP are encoded interdependently using motion estimation, while the frames belonging to different GOPs are encoded independently. Note that the prediction-based coding schemes can lead to sophisticated dependencies between the video data. After being encoded, each GOP contains $N$ data units (DUs), each of which representing one or multiple encoded frames (e.g. one of I, P, and B frames)[3]. The attributes of the DUs are listed below.

*Size:* The size of DU $j \in \{1,\cdots,N\}$ is denoted as $l_j$ in packets[4], where $l_j \in [1, l_j^{\max}]$, and $l_j^{\max}$ is the maximum allowable size. The size of DU $j$ is the amount of packets when DU $j$ is generated by the video codec. For example, the size of DU $j$ at GOP $g$ is different from the size of DU $j$ at GOP $g+1$.

---

[2] Note that the video traffic can be generated in real-time or pre-encoded.

[3] In this paper, we consider that the video is encoded using a fixed GOP structure. However, the proposed traffic representation can also be adapted to the dynamic GOP structure.

[4] For simplicity, we assume in this paper that each packet has the same length, but this does not affect our proposed solution. It just simplifies our exposition given the space limitations.

To simplify the exposition, this is modelled as an i.i.d. random variable[5].

*Distortion impact*: Each DU $j$ has a distortion impact $q_j$ per packets, which is assumed to be constant for all the GOPs.

*Delay deadline*: We define the relative delay deadline (RDD) of DU $j$ as the difference between the delay deadlines of DU 1 and DU $j$ in the same GOP and is denoted by $d_j$ (measured in time slots). Hence, $d_1 = 0$ and $d_j \leq T$. If the delay deadline for DU 1 at the first GOP is set to be $d_0$, then the delay deadline of DU $j$ at the $g$-th GOP is $d_0 + d_j + (g-1)T$.

*Dependency*: The dependencies between the DUs within one GOP are expressed as a directed acyclic graph (DAG). The DAG remains the same for a fixed GOP structure. One illustrative example of DAGs for video data is given in Figure 1. In this paper, we assume that, if DU $j$ depends on DU $j'$ (i.e. there exists a path directed from DU $j$ to DU $j'$ and denoted by $j \prec j'$.), then $d_j \geq d_{j'}$ and $q_j \leq q_{j'}$. In other words, DU $j'$ should be decoded prior to DU $j$ and DU $j'$ has higher distortion impact.

### *B. Traffic state representation at each time slot*

In this subsection, we model the video traffic which can be potentially transmitted at each time slot. At time slot $t$, as in [2], we assume that the wireless user will only consider for transmission DUs with delay deadlines in the range of $[t, t+W)$, where $W$ is referred to as the scheduling time window (STW) and assumed to be given a priori[6]. In this paper, we further assume that STW is chosen to satisfy the following condition: if DU $j$ directly depends on DU $j'$ (i.e. there is a direct arc from $j$ to $j'$), then $d_j - d_{j'} < W$. This assumption ensures that DU $j$ and $j'$ can be in one STW.

We define the traffic state as the DUs within the STW at time slot $t$, which is denoted by $\mathcal{T}_t = (\mathcal{G}_t, \mathcal{B}_t)$. In the traffic state $\mathcal{T}_t$, $\mathcal{G}_t = \left\{ j_g \mid \exists g \in \mathbb{N}, d_0 + d_{j_g} + (g-1)T \in [t, t+W), \text{Dependencies of } j_g \text{ expressed by } DAG \right\}$ is called the dependency pattern of DUs within the STW. The dependency pattern $\mathcal{G}_t$ represents the number of DUs that can be potentially transmitted during the next time slot and the dependencies between them. In the example illustrated in Figure 1, $\mathcal{G}_t = \{1_g, 2_g, 3_g \mid 1_g \prec 2_g \prec 3_g, 1_g \prec 3_g\}$ and $\mathcal{G}_{t+1} = \{4_g, 5_g, 1_{g+1} \mid 4_g \prec 5_g\}$, where $g$ is the GOP index. $\mathcal{B}_t = \{b_j \mid j \in \mathcal{G}_t\}$ represents the amount of packets at each DU available for potential transmission at time slot $t$. Note that $b_j \le l_j$. From the example in Figure 1, we note that the transition of $\mathcal{G}_t$ is deterministic and periodic for a predetermined GOP structure and hence, it is Markovian and denoted by $p_{\mathcal{G}}(\mathcal{G}' \mid \mathcal{G})$. The traffic state $\mathcal{T}_t$ is able to capture heterogeneous video traffic and is a super-set of existing well-known single-buffer models [9][10] (i.e. which ignore both the packet dependencies and delay deadlines) or multi-buffer models [3][12][17][35] (i.e. which ignore the packet dependencies or the delay deadlines).

[5] The DU size can also be modeled as a random variable depending on the previous DUs [28].

[6] Note that STW can be determined based on the channel conditions experienced by the user at each time slot. For example, the wireless user may set the STW to be small when the channel conditions are poor, and the STW to be large whenever the channel condition are good.

### *C. Scheduling policy*

Given a transmission rate [7] $r_t$, the wireless user has to determine the amount of data to be transmitted for each DU in $\mathcal{G}_t$, through its scheduling policy. The scheduling policy maps the current traffic state $\mathcal{T}_t$ and transmission rate $r_t$ into the amount of packets transmitted during each DU, $\boldsymbol{y}_t = [y_j \mid j \in \mathcal{G}_t]$ in the current STW, i.e. $\pi(\mathcal{T}_t, r_t) = \boldsymbol{y}_t$. Formally, the scheduling policy $\pi$ satisfies the following conditions[8]: (i) Underflow constraint: $0 \le y_j \le b_j,\ j \in \mathcal{G}$; (ii) Rate constraint: $\sum_{j \in \mathcal{U}_t} y_j \le r_t$. The set of possible policies in each traffic state $\mathcal{T}_t$ given a certain transmission rate $r_t$ is denoted by $\mathcal{P}(\mathcal{T}_t, r_t)$.

### *D. Traffic state transition and immediate reward*

In the following, we discuss the transition of the traffic state $\mathcal{T}_t$, given the transmission rate $r_t$. First, the transition of the dependency pattern $\mathcal{G}_t$ is $p_{\mathcal{G}}(\mathcal{G}_{t+1} \mid \mathcal{G}_t)$, which is deterministic and does not depend on the transmission rate $r_t$. In order to compute the transition from $\mathcal{B}_t$ to $\mathcal{B}_{t+1}$, we separate $\mathcal{G}_{t+1}$ into two disjoint sets: $\mathcal{G}_t \cap \mathcal{G}_{t+1}$ and $\mathcal{G}_{t+1} / \mathcal{G}_t$ [9]. It is clear that $\mathcal{G}_{t+1} = (\mathcal{G}_t \cap \mathcal{G}_{t+1}) \cup (\mathcal{G}_{t+1} / \mathcal{G}_t)$.

First, we consider the DUs in the set of $\mathcal{G}_t / \mathcal{G}_{t+1}$ that will expire before time slot $t+1$. In this paper,

---

[7] The transmission rate can be determined by the allocated network resource and transmission strategies at the layers below the application layer

[8] Similar constraints are also considered in [29]. However, the authors therein did not consider the time-varying transmission rate and foresighted packet scheduling decisions aimed at maximizing the long-term video quality.

[9] Here $\mathcal{G}_{t+1} / \mathcal{G}_t = \mathcal{G}_{t+1} - \mathcal{G}_{t+1} \cap \mathcal{G}_t$.

we consider the case that, if DU $j \in \mathcal{G}_t / \mathcal{G}_{t+1}$ has the remaining data $b_j - y_j$ greater than a certain threshold[10] (say $V_j$, i.e. $b_j - y_j > V_j$), then all its descendants will be undecodable (or useless) and hence, will be discarded for transmission. Let us denote $\mathcal{E}_t = \{j \mid j \in \mathcal{G}_t / \mathcal{G}_{t+1}, b_j - y_j \geq V_j\}$ to be the set of DUs that will expire before time slot $t+1$ and will also result in unusable descendant DUs due to large amount of data lost. We further denote $\epsilon_t = \{j \mid j \in \mathcal{G}_t \cap \mathcal{G}_{t+1}, b_j = -1\}$ to be the set of DUs which are in $\mathcal{G}_t \cap \mathcal{G}_{t+1}$, but have $b_j = -1$ (we use -1 to indicate that a certain DU is useless.). Then, the transition from $\mathcal{B}_t$ to $\mathcal{B}_{t+1}$ is computed as:

$$b_j \leftarrow \begin{cases} -1 & \text{if } \mathcal{A}(j) \cap (\mathcal{E}_t \cup \epsilon_t) \neq \varnothing \\ b_j - y_j & \text{if } j \in \mathcal{G}_t \cap \mathcal{G}_{t+1} \text{ and } \mathcal{A}(j) \cap (\mathcal{E}_t \cup \epsilon_t) = \varnothing \\ l_j & \text{if } j \in \mathcal{G}_{t+1} / \mathcal{G}_t \text{ and } \mathcal{A}(j) \cap (\mathcal{E}_t \cup \epsilon_t) = \varnothing \end{cases} \tag{1}$$

Note that $\mathcal{A}(j)$ represents all the ancestors of DU $j$, including itself. The first line in Eq. (1) means that DU $j$ is unusable, because its ancestors are not successfully received. The second line means that DU $j$ is inherited from the previous traffic state, i.e. DU $j$ did not expire at the current moment. The third line means that DU $j$ is a new DU entering the STW. Since the transition of the dependency pattern is deterministic and the incoming DUs are i.i.d., it is clear that the transition of $\mathcal{B}_t$ is Markovian as well and hence, the traffic state transition is Markovian. The transition probability from $\mathcal{B}_t$ to $\mathcal{B}_{t+1}$ is denoted as $p_{\mathcal{B}}(\mathcal{B}_{t+1} \mid \mathcal{B}_t, \mathcal{G}_t, \mathcal{G}_{t+1}, \boldsymbol{y}_t, r_t)$, and depends on the dependency pattern transition, scheduling policy and the available transmission rate. The traffic state transition can be rewritten as $p_{\mathcal{T}}(\mathcal{T}_{t+1} \mid \mathcal{T}_t, \boldsymbol{y}_t, r_t) = p_{\mathcal{G}}(\mathcal{G}_{t+1} \mid \mathcal{G}_t) p_{\mathcal{B}}(\mathcal{B}_{t+1} \mid \mathcal{B}_t, \mathcal{G}_t, \mathcal{G}_{t+1}, \boldsymbol{y}_t,, r_t)$. Given the scheduling policy $\boldsymbol{y}_t = \pi(\mathcal{T}_t, r_t)$ and transmission rate $r_t$, the distortion reduction experienced by the wireless user can be computed as

$$u_t(\mathcal{T}_t, \boldsymbol{y}_t, r_t) = \sum_{j \in \mathcal{U}_t} q_j \cdot y_j \,. \tag{2}$$

[10] More complicated models can also be developed for the dependency impact between two DUs. For example, an error propagation function can be defined to capture the dependency impact. However, in this case, the error propagation function should be included into the traffic state which may dramatically increase the state space. In this paper, we only consider a simple dependency impact model (i.e. on-off model).

## III. DYNAMIC OPTIMIZATION FOR A SINGLE USER

In this section, we first consider the optimization of both the packet scheduling and resource acquisition for a single wireless video user experiencing a slow fading wireless channel. In each time slot, the wireless user experiences a channel condition $h_t$. We assume that the channel condition $h_t$ remains constant within one time slot, but varies across time slots. The changes of $h_t$ can be modelled as a finite state Markov chain (FSMC) [24] with the state transition probability given by $p_h(h_{t+1} \mid h_t)$, which is independent of the traffic state transition. The transmission rate attained by the wireless user is determined by $r_t(h_t, x_t)$, where $x_t \in X$ represents the amount of network resource (e.g. the transmission time in the TDMA-like network [25] as discussed in Section IV) acquired by the wireless user from the network and $X$ represents the set of possible resource allocations. As we will discuss in Section IV for the multi-user video transmission, the resource acquisition will be affected by other users. The transmission rate function $r_t(h_t, x_t)$ is assumed to be an increasing function of $x_t$, given the channel condition $h_t$ [36], and thereby, a larger $x_t$ leads to a higher transmission rate $r_t$.

We define the state for the wireless user at time slot $t$ as $s_t = (\mathcal{T}_t, h_t) \in S$, which includes the video traffic state and channel state, which satisfies the Markovian property since both the traffic state and channel state are Markovian. Then, the wireless user state transition is expressed by

$$p(s_{t+1} \mid s_t, \boldsymbol{y}_t, x_t) = p_{\mathcal{T}}(\mathcal{T}_{t+1} \mid \mathcal{T}_t, \boldsymbol{y}_t, r_t) p_h(h_{t+1} \mid h_t). \tag{3}$$

At each state $s_t$, the wireless user takes the actions including the resource acquisition $x_t$ and scheduling $\boldsymbol{y}_t$, thereby leading to the immediate utility $u_t(s_t, \boldsymbol{y}_t, x_t) - \lambda_{s_t} x_t$, where $\lambda_{s_t}$ is interpreted as the resource price as in [2]. Note that we express $u_t(\mathcal{T}_t, \boldsymbol{y}_t, r_t(h_t, x_t))$ as $u_t(s_t, \boldsymbol{y}_t, x_t)$ to emphasize that the immediate utility is a function of the state $s_t$, scheduling action $\boldsymbol{y}_t^i$ and allocated time $x_t^i$.

In this section, we assume that $\lambda_s$ is determined a priori. In Section V, we will discuss how the resource price can be determined in a multi-user scenario. The objective of the wireless user is to maximize its expected discounted accumulated utility (we call this “single-user primary problem (SUP)”,)

as follows[11]:

$$\max_{\substack{\boldsymbol{y}_t \in \mathcal{P}(s_t,x_t) \\ x_t \geq 0, t \geq 0}} \sum_{s_0 \in S} v(s_0) E\left\{ \sum_{t=0}^{\infty} \alpha^t \left( u_t\left(s_t, \boldsymbol{y}_t, x_t\right) - \lambda_{s_t} x_t \right) \mid s_0 \right\}, \tag{SUP}$$

where $\alpha$ is the discounted factor in the range[12] of $[0,1)$, and $v(s_0)$ is the distribution of the initial state. The reasons why we consider the discounted accumulated utility are: (1) for our considered delay-sensitive applications, the data needs to be sent out as soon as possible to avoid missing delay deadlines; and (2) since a wireless user may encounter unexpected environmental dynamics in the future, it may care more about its immediate reward rather than the future reward. Based on the discussion in Section II.D, the transition of the state $s_t$ only depends on the current actions $x_t$ and $\boldsymbol{y}_t$. Hence, the problem above can be formulated as a MDP.

Note that unlike the previous video transmission solutions in [1]-[7], here we explicitly take into consideration the heterogeneous characteristics of video traffic (represented by the traffic states) and time-varying channel conditions (represented as channel states). Similar to the work in [2], we optimize the trade-off between the consumed resource and the received reward in terms of distortion reduction, but focusing on a dynamic setting. The optimization of SUP is called the foresighted optimization for video transmission since it considers the impact of current decisions on the future utility. Based on [19], the optimization of SUP can be solved using the following Bellman's equations:

$$U(s,\boldsymbol{\lambda}) = \max_{\substack{\boldsymbol{y} \in \mathcal{P}(s,x) \\ x \geq 0}} \left[ u(s,\boldsymbol{y},x) - \lambda_s x + \alpha p(s' \mid s,\boldsymbol{y},x) U(s',\boldsymbol{\lambda}) \right], \tag{4}$$

where $U(s,\boldsymbol{\lambda})$ is the optimal reward-to-go starting at state $s$, given $\boldsymbol{\lambda} = [\lambda_s]_{s \in S}$. The Bellman's equations can be solved using the value iteration or policy iteration methods. One solution to solve the SUP is proposed in our previous work [35].

# IV. Multi-user Wireless Video Transmission Formulation

In Section III, we have formulated the dynamic optimization problem for the single-user video transmission. In this section, we aim to formulate the problem of multi-user video transmission over a slow fading wireless channel. We will show that the single-user video transmission serves as a subproblem of the multi-user problem in Section V, which is illustrated in Figure 2.

---

[11] In this formulation, we interchangeably express the admissible policy as $\mathcal{P}(\mathcal{T}_t, r_t)$ and $\mathcal{P}(s_t, x_t)$.

[12] Our solutions discussed below are also applicable to the problem of maximizing the average accumulated utility by allowing $\alpha \to 1$.

The users are indexed by $i \in \{1, \cdots, M\}$, where $M$ is the number of users sharing the channel. At each time slot $t$, the channel condition experienced by user $i$ is represented by $h_t^i$ (the superscript $i$ represents user $i$ and the same in the below), and the transition of $h_t^i$ is independent of the transitions of other users' channel conditions.

Recall that in this paper we assumed that the multiple users access the shared channel using the TDMA-like protocol [25]. Then, at each time slot, the portion of time allocated to user $i$ is denoted by $x_t^i \in [0,1]$. Due to the resource constraint, the allocations to all the users satisfy the following inequality:

$$\sum_{i=1}^{M} x_t^i \leq 1, \forall t \,. \tag{5}$$

In this paper, we consider a collaborative multi-user video transmission problem with the goal of maximizing the expected discounted accumulated video quality of all the users under the stage resource constraints (we call this problem "the multi-user primary problem with stage resource constraints - MUP/SRC"), i.e.

$$\begin{gathered} U^* = \max_{\boldsymbol{y}_t^i, x_t^i, i=1,\cdots,M} \sum_{s_0^1 \in S^1, \cdots, s_0^M \in S^M} \prod_{i=1}^{M} v\left(s_0^i\right) E\left[\sum_{t=0}^{\infty} \sum_{i=1}^{M} (\alpha)^t u_t^i\left(s_t^i, \boldsymbol{y}_t^i, x_t^i\right) \mid s_0^1, \cdots, s_0^M\right] \\ s.t.\ \boldsymbol{y}_t^i \in \mathcal{P}^i\left(s_t^i, x_t^i\right),\ \sum_{i=1}^{M} x_t^i \leq 1, \forall t \geq 0 \end{gathered} \tag{MUP/SRC}$$

where we assume that the initial states of the video users are independent. Note that, in this paper, we consider that each user has the same discounted factor[13].

Based on the discussion on the video traffic representation in Section II and the formulation for the single-user video transmission, the multi-user transmission problem in MUP/SRC can be also formulated as an MUMDP. Specifically, we define the state of the multi-user system as $\boldsymbol{s} = \left(s^1, \cdots, s^M\right)$. The action performed by each user is $\boldsymbol{a} = \left(\left(x^1, \boldsymbol{y}^1\right), \cdots, \left(x^M, \boldsymbol{y}^M\right)\right)$. It is easy to verify that $p(\boldsymbol{s}' \mid \boldsymbol{s},\boldsymbol{a}) = \prod_{i=1}^{M} p^i\left(s^{i\prime} \mid s^i,\boldsymbol{y}^i,x^i\right)$, given the resource allocation $x^i,\forall i$. The reward at each time slot is given by $u_t = \sum_{i=1}^{M} u_t^i\left(s_t^i,\boldsymbol{y}_t^i,x_t^i\right)$.

---

[13] Since we consider a collaborative multi-user video transmission, for simplicity, we enforce that each user has the same discounted factor. However, in general, different users may have different discounted factor, especially in the non-collaborative scenarios.

We note that, when $\alpha = 0$ (i.e. all the users make the myopic decision), the MUMDP problem reduces to the traditional multi-user NUM-based resource allocation problems for video transmission [13]~[18]:

$$\begin{aligned} &\max \sum_{i=1}^{M} u^i\left(s^i,\boldsymbol{y}^i,x^i\right) \\ &s.t.\ \boldsymbol{y}^i \in \mathcal{P}^i\left(s^i,x^i\right),\quad \sum_{i=1}^{M} x^i \le 1 \end{aligned} . \tag{6}$$

However, we consider here the dynamic optimization for the multi-user video transmission by taking into account the resource allocation and corresponding scheduling across time (i.e. $\alpha \neq 0$).

From [19], we know that, for this multi-user MDP problem, there is at least one optimal stationary policy that only depends on the current multi-user system state. Hence, in this paper, we restrict our focus to the stationary policies, i.e. the policy only depends on the current state. Then, solving the maximization problem in MUP/SRC is equivalent to solving the following Bellman's equations [19]:

$$U(\boldsymbol{s}) = \max_{\substack{\boldsymbol{y}^i \in \mathcal{P}^i\left(s^i,x^i\right), i=1,\cdots,M \\ \sum_{i=1}^{M} x^i \le 1}} \left[\sum_{i=1}^{M} u_i\left(s^i,\boldsymbol{y}^i,x^i\right) + \alpha \sum_{\boldsymbol{s}'} \prod_{k=1}^{M} p\left(s^{k\prime} \mid s^k,\boldsymbol{y}^k,x^k\right) U(\boldsymbol{s}')\right], \forall \boldsymbol{s}, \tag{7}$$

and $U^* = \sum_{s_0^1 \in S^1,\cdots,s_0^M \in S^M} \prod_{i=1}^{M} v\left(s_0^i\right)\ U(\boldsymbol{s}_0)$.

From this Bellman's equations, we have the following observations:

- To solve the Bellman's equations, we can use the centralized value iteration or policy iteration to find the optimal reward-to-go $U(\boldsymbol{s})$ for the multi-user MDP problem [19]. However, this centralized solution requires knowing all the users' information (state spaces, action spaces, transition probabilities, and utility functions) and also has a high computation complexity. Hence, this centralized solution is not applicable to the multi-user wireless video transmission.
- The coupling among the multiple users' video transmission is only through the resource allocation

performed at each time slot. For example, the optimal scheduling policy performed by each user $i$ depends on the multi-user system state through the resource allocation $x^i$. Then, given the resource allocation $x^i$, the scheduling policy is independent of other users' states. This type of MUMDP is referred as the weakly-coupled MDP [20] and the decomposition into multiple local MDPs is possible.

In the next section, we will discuss how the multi-user MDP problem can be decomposed when the resource allocation is dynamic and depends on the multi-user system's state. The relationships among proposed solutions are illustrated in Figure 2.

## V. DUAL DECOMPOSITION OF MULTI-USER MDP

In this section, we will consider the dual problem of the multi-user MDP by relaxing the per-stage resource constraints and show how we can decompose the MUMDP. First, in Subsection A, we introduce a per-state Lagrangian multiplier associated with the resource constraint at each state. This dual solution leads to the zero duality compared to the primary problem MUP/SRC, but requires a centralized solution since the resource price depends on multi-user state which cannot be observed by each individual user. Then, in Subsection B, we impose a uniform resource price, which is independent of the multi-user state. With this uniform resource price, the MUMDP problem can be decomposed into multiple local MDPs, which represent a dynamic cross-layer optimization problem that can be separately solved by each individual user. This decomposition is promising since (i) it enables each user to optimize its packet scheduling and resource acquisition independently of other users; and (2) the network coordinator only needs to simply update the resource price, which involves only very few computations.

### A. *Dual solution with per-state resource prices*

At each state $\boldsymbol{s}_t$, we introduce a Lagrangian multiplier $\lambda_{\boldsymbol{s}_t}$ associated with the resource constraint $\left(\sum_{i=1}^{M} x^i - 1\right)$ at each state $\boldsymbol{s}_t$. Then the dual function is given by

$$U(\boldsymbol{\lambda}) = \max_{\substack{\boldsymbol{y}_t^i \in \mathcal{P}^i(s_t^i, x_t^i), x_t^i \geq 0, \\ i=1,\cdots,M, \\ t \geq 0}} \sum_{s_0^1 \in S^1, \cdots, s_0^M \in S^M} \prod_{i=1}^{M} v(s_0^i) E\left[\sum_{t=0}^{\infty} (\alpha)^t \sum_{i=1}^{M} \left( u_t^i(s_t^i, \boldsymbol{y}_t^i, x_t^i) - \lambda_{\boldsymbol{s}_t} x_t^i + \frac{\lambda_{\boldsymbol{s}_t}}{M} \right) \mid \boldsymbol{s}_0 \right], \quad (8)$$

with $\boldsymbol{\lambda} = [\lambda_{\boldsymbol{s}}]$. As in Section III, we can interpret the Lagrangian multiplier $\lambda_{\boldsymbol{s}}$ as the resource price in state $\boldsymbol{s}$. We refer to this as "pre-state resource price". Then, $\lambda_{\boldsymbol{s}} x^i$ is the cost user $i$ has to pay in state $\boldsymbol{s}$ and $\lambda_{\boldsymbol{s}} \cdot 1$ is the amount of revenue received by the multi-user system by allowing the users to consume the resources (i.e. access the wireless channel). However, we should note that, in our collaborative communications, the resource price is used in order to efficiently allocate the limited resource, instead of maximizing the revenue of the multi-user system.

The multi-user dual problem with the per-state resource price (MUD/PSRP) is then given by

$$U^{\boldsymbol{\lambda},*} = \min_{\boldsymbol{\lambda} \geq 0} U(\boldsymbol{\lambda}). \qquad \text{(MUD/PSRP)}$$

The following proposition proves that the dual problem MUD/PSRP has zero duality gap compared to the primary problem in MUP/SRC and thus, the optimal time allocation and scheduling policies corresponding to the optimal resource price $\lambda_{\boldsymbol{s}}$ at each state are also the optimal policies to the primary problem.

**Proposition 1**: $U^{\boldsymbol{\lambda},*} = U^*$.

*Proof*: Due to the limited space, we omit the proof of this proposition. However, the proof can be easily performed by showing that the optimal scheduling policy leads to an objective function which is a concave function. This is because, in each time slot, the optimal scheduling policy always transmits the packets resulting in the highest increase in the long-term utility. Given the concavity of the objective function and the compactness of the feasible resource allocation space, we can prove that the duality gap is zero [26].

Similar to the Bellman's equations in Eq. (7) for the primary problem MUP/SRC, we have the following Bellman's equations corresponding to the dual function in MUD/PSRP:

$$U(\boldsymbol{s},\boldsymbol{\lambda}) = \max_{\substack{\boldsymbol{y}^i \in Y^i \\ x^i \geq 0 \\ i=1,\cdots,M}} \left[ \sum_{i=1}^{M} \left( u_i\left(s^i, \boldsymbol{y}^i, x^i\right) - \lambda_{\boldsymbol{s}} x^i + \frac{1}{M}\lambda_{\boldsymbol{s}} \right) + \alpha \sum_{\boldsymbol{s}'} \prod_{k=1}^{M} p\left(s^{k\prime} \mid s^k, \boldsymbol{y}^k, x^k\right) U(\boldsymbol{s},\boldsymbol{\lambda}) \right], \forall \boldsymbol{s} \quad (9)$$

We note that, by setting $\alpha = 0$, the Bellman's equations above degrade to the dual solutions [13][15]

to the conventional multi-user video transmission as shown in Eq. (6). The degraded Bellman's equations can be decomposed into multiple sub-equations, each corresponding to one user, by letting the user know the resource price. However, in general, this Bellman's equation cannot be decomposed into independent subproblems which can be autonomously solved by each user, since the Bellman's equations are coupled through the resource price $\lambda_{\boldsymbol{s}}$, which varies with the state of the multi-user system. Hence, a centralized solution has to be deployed by the network coordinator, which requires all the users' information, as in the primary solution to MUP/SRC.

*B. Dual solution with uniform resource price*

In Subsection A, we assumed that, depending on the state of the multi-user system, a different resource price is determined. However, the drawback of this is that the Bellman's equations in Eq. (9) cannot be decomposed, thereby requiring a centralized solution. In this subsection, we consider a scenario where the same price (referred to as "uniform resource price") is imposed in all the states of the multi-user system, i.e. $\lambda_{\boldsymbol{s}} = \lambda, \forall \boldsymbol{s}$. Then, the dual function is given by

$$U(\lambda) = \max_{\substack{\boldsymbol{y}_t^i \in \mathcal{P}^i(s_t^i, x_t^i), x_t^i \geq 0, \\ i=1,\cdots,M, \\ t \geq 0}} \sum_{s_0^1 \in S^1, \cdots, s_0^M \in S^M} \prod_{i=1}^{M} v(s_0^i) E\left[\sum_{t=0}^{\infty} (\alpha)^t \sum_{i=1}^{M} \left( u_t^i(s_t^i, \boldsymbol{y}_t^i, x_t^i) - \lambda x_t^i + \frac{\lambda}{M} \right) \mid \boldsymbol{s}_0 \right]. \quad (10)$$

By minimizing over the uniform resource price $\lambda$, we have the multi-user dual problem with uniform resource price (MUD/URP):

$$U^{\lambda,*} = \min_{\lambda \geq 0} U(\lambda) \qquad \text{(MUD/URP)}$$

Interestingly, by setting the uniform resource price, the dual problem MUD/URP is not dual to the primary problem in MUP/SRC. Instead, it is dual to the following problem:

$$\begin{aligned} \hat{U}^* = \max_{\boldsymbol{y}_t^i, x_t^i, i=1,\cdots,M} & \sum_{s_0^1 \in S^1, \cdots, s_0^M \in S^M} \prod_{i=1}^{M} v(s_0^i) E\left[\sum_{t=0}^{\infty} \sum_{i=1}^{M} (\alpha)^t u_t^i(s_t^i, \boldsymbol{y}_t^i, x_t^i) \mid s_0^1, \cdots, s_0^M \right] \\ s.t.\ & \boldsymbol{y}_t^i \in \mathcal{P}^i(s_t^i, x_t^i), \ \sum_{t=0}^{\infty} \sum_{i=1}^{M} (\alpha)^t \left( x_t^i - \frac{1}{M} \right) \leq 0 \end{aligned} \qquad \text{(MUP/ARC)}$$

We call this optimization – "the multi-user video transmission optimization with accumulated resource constraint (MUP/ARC)". The duality between MUD/URP and MUP/ARC can be easily verified. Similar to Proposition 1, we can prove that the duality gap between MUD/URP and MUP/ARC is zero.

We further notice that the resource constraint in the primary problem MUP/SRC satisfies the following condition:

$$\left\{x_t^i, i = 1,\cdots,M, t \geq 0 \mid \sum_{i=1}^{M} x_t^i \leq 1\right\} \subset \left\{x_t^i, i = 1,\cdots,M, t \geq 0 \mid \sum_{t=0}^{\infty}\sum_{i=1}^{M} (\alpha)^t \left(x_t^i - \frac{1}{M}\right) \leq 0\right\}, \quad (11)$$

which means that the feasible resource allocations in the MUP/SRC is a subset of the feasible resource allocations in the MUP/ARC. Then, comparing to the dual solution with state-wise prices, we have the following proposition which shows that $U^{\lambda,*}$ serves as an upper bound of the optimal value for the primary problem.

**Proposition 2**: $U^{\lambda,*} = \hat{U}^* \geq U^* = U^{\boldsymbol{\lambda},*}$.

This proof is based on the fact that there is no duality gap in the two primary-dual problems (i.e. MUP/SRC and MUD/PSRP, MUP/ARC and MUD/URP) and that the feasible resource allocations in MUP/SRC is a subset of the feasible allocations in MUP/ARC as shown in Eq. (11). The details of the proof are omitted due to the limited space.

The Bellman's equations corresponds to the dual function in Eq. (10) can be decomposed into $M$ local Bellman's equation, each corresponding to one user, which is presented in the following theorem.

**Theorem 2**: Given $\lambda_{\boldsymbol{s}} = \lambda, \forall \boldsymbol{s}$, the optimization in Eq. (10) is given by

$$U(\lambda) = \sum_{i=1}^{M} \sum_{s_0^i} v\left(s_0^i\right) U^i\left(s_0^i, \lambda\right), \quad (12)$$

with $U^i\left(s^i, \lambda\right)$ satisfying the following local Bellman's equation:

$$U^i\left(s^i, \lambda\right) = \max_{y^i, x^i}\left[u^i\left(s^i, y^i, x^i\right) - \lambda x^i + \frac{1}{M}\lambda + \alpha \sum_{s^{i\prime}} p\left(s^{i\prime} \mid s^i, \boldsymbol{y}^i, x^i\right) U^i\left(s^{i\prime}, \lambda\right)\right] \quad (13)$$

Proof: The key idea to prove this is that, by introducing the uniform resource price, the utility functions and state transition probabilities of all wireless users are separable which makes the reward-to-go functions separable. The details can be seen in Appendix A.

The key result of Theorem 2 is that $U(\boldsymbol{s}, \lambda)$ can be decomposed into $M$ local Bellman's equations, which can be solved in a distributed fashion. Each user can solve its own Bellman's equations (and accordingly solve its own cross-layer optimization problem) provided the resource price $\lambda$. This local

Bellman's equations correspond to the local MDP discussed in Section III.

Next, we discuss how the resource price can be updated. Given the resource price $\lambda$, each user can solve its own Bellman's equations using e.g. value iteration, which results in the optimal resource allocation $x^{i,*}\left(s^i,\lambda\right)$ and scheduling policy $\boldsymbol{y}^{i,*}\left(s^i,\lambda\right)$. Note that the resource acquisition is independent of other users' state given the uniform resource price. In the following proposition, we formally compute the subgradient with respect to the resource price $\lambda$ for the dual problem MUD/URP, which will be used to update the resource price in each iteration.

**Proposition 3**: The subgradient with respect to $\lambda$ is given by

$$\sum_{i=1}^{M} Z^i - \frac{1}{1-\alpha}, \tag{14}$$

where $Z^i = \sum_{s_0^i} v\left(s_0^i\right)\boldsymbol{e}_{s_0^i}^T\left(I - P^i\right)^{-1}\boldsymbol{x}^i\left(\lambda\right)$ is the expected discounted accumulated resource consumption (note that the expectation is taken over all the possible sample paths), and $P^i$ is the state transition probability matrix, and $\boldsymbol{e}_{s^i}$ is the vector with the $s^i$ component being 1 and others being zero.

*Proof*: The key idea is to show $\sum_{i=1}^{M} Z^i - \frac{1}{1-\alpha}$ satisfies the subgradient definition [26]. The details can be found in Appendix B.

Using the subgradient method, the resource price is then updated as follows:

$$\lambda^{k+1} = \left[\lambda^k + \beta^k\left(\sum_{i=1}^{M} Z^i - \frac{1}{1-\alpha}\right)\right]^+ \tag{15}$$

We notice that the subgradient computed in Eq. (14) accounts for not only the current resource constraint, but also the future constraints since MUMDP aims to maximize the long-term utility. The subgradient method shown in Eq. (15) converges to the optimal dual solution due to the concavity of the objective function. The advantages of the decomposition developed above for multi-user video transmission are summarized as follows:

- Given a uniform resource price, each wireless user can solve its own local MDP independently of other users. This enables us to decompose the consider multi-user video transmission problem by enabling each user to autonomously optimize its the packet scheduling and resource acquisition.

- This decomposition allows the network coordinator to simply update the scalar resource price as shown in Eq. (15). Furthermore, the proposed approach only requires two scalar messages[14] (as shown in Figure 3) to be exchanged between the wireless users and the network coordinator. This significantly simplifies the design of the network coordinator (e.g. access points) and reduces the cost of building a wireless network to support video applications.

- Unlike the dual solution to the NUM-based multi-user video transmission [15], in which the network coordinator has to find the optimal resource allocation to all the users before each user optimizes the packet scheduling, our approach allows the resource allocation and packet scheduling optimization to be performed simultaneously. This significantly reduces the incurred delay, since each user does not need to wait for the optimal resource allocation.

Previously, we enforced a uniform price for all the states, which enables a decomposition in the dual function computation and provides an upper bound on the utility function $U^*$ (Recall that $U^{\lambda,*} \geq U^*$). However, the solution to the dual problem may be infeasible (i.e. violating the resource constraint in each time slot). We note that the optimal allocation can be obtained by solving a one-stage multi-user resource allocation problem (e.g. as in [20]) in each time slot. However, this again requires high computational and communication complexity. In order to avoid the resource constraints being violated, i.e. $\sum_{i=1}^{M} x^{i,\lambda^*}\left(s^i\right) > 1$, we proportionally scale down the resource allocation as follows:

$$\hat{x}^{i,\lambda^*}(\boldsymbol{s}) = \frac{x^{i,\lambda^*}\left(s^i\right)}{\sum_{j=1}^{M} x^{j,\lambda^*}\left(s^j\right)}, \tag{16}$$

in order to satisfy the resource constraint, i.e. $\sum_{i=1}^{M} \hat{x}^{i,\lambda^*}\left(s^i\right) = 1$. This scaling can be performed by the network coordinator: at the beginning of each time slot, the users submit the required resource $x^{i,\lambda^*}\left(s^i\right)$ to the network coordinator, and the coordinator performs the resource allocation scaling, if the resource constraint is violated. After the scaling, the network coordinator polls the users according to the scaled

[14] In protocol design, other messages like the hand-shaking messages are needed for the successful transmission. However, we ignore this type of messages since they are independent of our proposed algorithms.

resource allocation [25].

## VI. Distributed Modified Actor-Critic Learning

In Section V, we have discussed how the MUMDP can be decomposed such that each user can autonomously determine its optimal transmission strategy, using the uniform resource price. However, when implementing this solution in practice, we still face the following difficulties: (i) in the distributed solution, each user still has to solve its own local MDP problem for each updated resource price, which still leads to a very high computation complexity for each user; (ii) the channel states and incoming DUs dynamics are often difficult to characterize a priori, such that the single-user MDP cannot be solved online; (iii) without being able to optimally solve the local MDP, the subgradient in Eq. (14) cannot be computed. To address these challenges, we aim in this section at developing an online learning algorithm. Specifically, we deploy a modified actor-critic learning algorithm [27] to solve the single-user MDP online and the stochastic subgradient method [26] to update the uniform resource price. The advantages of the online learning are: at each time slot, the wireless user only needs to perform limited computations (i.e. incurs a low computation complexity); and, the online learning does not require a priori knowledge of the channel states and incoming DUs dynamics.

### A. Updating state-value function and resource acquisition policy

To perform the actor-critic learning algorithm for multi-user video transmission, as shown in Figure 5, each wireless user needs to implement two components: the actor and the critic. The actor is endowed with a resource acquisition policy representation $\rho^i\left(s^i,x^i,\lambda\right)\in\mathbb{R}_+$, which indicates the tendency to select the resource acquisition action $x^i$ at the state $s^i$, i.e. the higher $\rho^i\left(s^i,x^i,\lambda\right)$ is, the larger the probability of selecting the action $x^i$ should be. The critic is endowed with the state value function $U^i\left(s^i,\lambda\right)$, which is used to evaluate the resource acquisition policy updated by the actor, and the higher $U^i\left(s^i,\lambda\right)$ is, the higher long-term utility the policy will provide. To evaluate the policy, the critic will keep updating the state value function, which is similar to the policy evaluation in the policy iteration algorithm [27]. At each time slot $t$, the wireless user $i$ selects a resource acquisition action $x_t^i$ based on the following

softmax method [27]:

$$\pi\left(s^{i}\right)=\sum_{x^{i}}\frac{e^{\rho\left(s^{i},x^{i}\right)}}{\sum_{x^{i\prime}}e^{\rho\left(s^{i},x^{i\prime}\right)}}x^{i}\ . \tag{17}$$

This resource acquisition $x_t^i$ is then submitted to the network coordinator which returns the true resource allocation $\hat{x}_t^i$ based on Eq. (16) (through polling [25]). With the scheduling policy $\tilde{\boldsymbol{y}}_t^i\left(s_t^i,\hat{x}_t^i\right)$ presented in Subsection B, the wireless user can receive the reward by the amount of $u_t\left(s_t^i,\tilde{\boldsymbol{y}}_t^i,\hat{x}_t^i\right)-\lambda\hat{x}_t^i$ and then moves to the next state $s_{t+1}^i$ by observing the channel condition $h_{t+1}^i$ and incoming data $\left[v_j \mid j\in\mathcal{G}_{t+1}/\mathcal{G}_t\right]$ . The conventional actor-critic learning algorithm performs the following two operations:

- state-value function update: $U^i\left(s_t^i,\lambda\right)\leftarrow U^i\left(s_t^i,\lambda\right)+\mu_t^i\left(\mathcal{G}_t,h_t^i\right)\delta_t^i\left(s_t^i,\hat{x}_t^i,\lambda\right)$; (18)
- resource acquisition policy update: $\rho\left(s_t^i,x_t^i,\lambda\right)\leftarrow\rho\left(s_t^i,x_t^i,\lambda\right)+\nu_t^i\left(\mathcal{G}_t,h_t^i,x_t^i\right)\delta_t^i\left(s_t^i,\hat{x}_t^i,\lambda\right)$, (19) where $\mu_t^i\left(\mathcal{G}_t,h_t^i\right)$ and $\nu_t^i\left(\mathcal{G}_t,h_t^i,x_t^i\right)$ are diminishing step-sizes and $\delta_t^i\left(s_t^i,x_t^i,\lambda\right)$ is the time-difference error which is computed as follows: $\delta_t^i\left(s_t^i,\hat{x}_t^i,\lambda\right)=u_t\left(s_t^i,\tilde{\boldsymbol{y}}_t^i,\hat{x}_t^i\right)-\lambda\hat{x}_t^i+\alpha U^i\left(s_{t+1}^i,\lambda\right)-U^i\left(s_t^i,\lambda\right)$.

In this paper, we separately update the scheduling policy and the resource acquisition policy which will allow us to update multiple states in one time slot instead of one state at one time slot as performed in the conventional on-line learning [27]. The details are discussed below. We first define the associated states, which are the states having the same dependency pattern and sharing the same channel condition as the current state. These states are denoted as $s^i\left(s_t^i\right)\in\left\{\left(\mathcal{T},h_t^i\right)\mid\mathcal{T}=\left(\mathcal{G}_t,\mathcal{B}\right),\forall\mathcal{B}\right\}$. Given the resource allocation $\hat{x}_t^i$ and the current channel condition $h_t^i$, as discussed in Subsection B, the scheduling policy can be computed independently of the next channel condition $h_{t+1}^i$ and incoming DUs $\{j\in\mathcal{G}_{t+1}/\mathcal{G}_t\}$. Assuming that the wireless user is now in the associated state $s^i\left(s_t^i\right)$ instead of $s_t^i$, we are again able to compute the scheduling policy $\tilde{\boldsymbol{y}}^i\left(s^i\left(s_t^i\right),\hat{x}_t^i\right)$ and virtually transmit the packets (i.e. not actually transmit them), which results in the virtual reward of $u_t^i\left(s^i,\tilde{\boldsymbol{y}}^i,\hat{x}_t^i\right)-\lambda\hat{x}_t^i$.

Besides being able to update the state-value function and resource acquisition policy only in the current state $s_t^i$, we are also able to update them in the associated state $s^i\left(s_t^i\right)$ using Eqs. (18) and (19), with the time-difference error computed as: $\delta^i\left(s^i\left(s_t^i\right),\hat{x}_t^i,\lambda\right)=u_t\left(s^i\left(s_t^i\right),\hat{\boldsymbol{y}}_t^i,\hat{x}_t^i\right)-\lambda\hat{x}_t^i+\gamma U^i\left(s'^i\left(s_t^i\right),\lambda\right)-U^i\left(s^i\left(s_t^i\right),\lambda\right)$, where $s'^i\left(s_t^i\right)$ is the next state transiting from the associated state $s^i\left(s_t^i\right)$.

The update of the state-value function and resource acquisition policy in all the associated states (including the current state) can significantly increase the learning performance (i.e. it increases faster the received distortion reduction) as compared to the standard on-line learning algorithm. In Section **Error! Reference source not found.**, we verify this through the concrete simulation.

*B. Greedy scheduling policy*

From the Bellman's equations shown in Eq. (4), we note that the optimal scheduling policy at each time slot can be computed as:

$$\tilde{\boldsymbol{y}}_t^i\left(s_t^i,\lambda\right)=\arg\max_{y\in\mathcal{P}\left(s_t^i,x_t^i\right)}\left[u_t\left(s_t^i,y,x_t^i\right)-\lambda x_t^i+\alpha\tilde{U}^i\left(\tilde{s}_t^i,\lambda\right)\right] \tag{20}$$

where $\tilde{s}_t^i$ is the post-decision state defined as $\tilde{s}_t^i=\left(\left(b_j-y_j\mid j\in\mathcal{G}_t\cap\mathcal{G}_{t+1}\right),h_t^i\right)$, which keeps the DUs which did not expire yet in $s_t^i$. $\tilde{U}\left(\tilde{s}_t^i,\lambda\right)$ is the average state-value function which is computed as

$$\tilde{U}^i\left(\tilde{s}_t^i,\lambda\right)=\sum_{h'\in\mathcal{H}}p_h\left(h'\mid h_t^i\right)\sum_{b_j,j\in\mathcal{G}'/\mathcal{G}}\prod_{j\in\mathcal{G}_{t+1}/\mathcal{G}_t}p\left(b_j\right)U^i\left(\left(\mathcal{T}',h'\right),\lambda\right), \tag{21}$$

where $\mathcal{T}'=\left[\left(b_j-y_j\mid j\in\mathcal{G}_t\cap\mathcal{G}_{t+1}\right),\left(b_j\mid j\in\mathcal{G}_{t+1}/\mathcal{G}_t\right)\right]$ is the next traffic state transiting from the post-decision state $\tilde{s}_t^i$. It is clear that $\tilde{U}^i\left(\tilde{s}_t^i,\lambda\right)$ does not depend on the channel state transition and incoming new DUs. Hence, given $\tilde{U}^i\left(\tilde{s}_t^i,\lambda\right)$, $\tilde{\boldsymbol{y}}_t^i\left(s_t^i,\lambda\right)$ can be computed independently of the next channel condition and incoming new DUs. However, since we do not know the distributions of channel state transition and incoming DUs, we cannot directly compute $\tilde{U}^i\left(\tilde{s}_t^i,\lambda\right)$ as in Eq. (21). Instead, we update $\tilde{U}^i\left(\tilde{s}_t^i,\lambda\right)$ as follows:

$$\tilde{U}^i\left(\tilde{s}_t^i,\lambda\right)\leftarrow\left(1-\varphi_t\left(\tilde{s}_t^i\right)\right)\tilde{U}^i\left(\tilde{s}_t^i,\lambda\right)+\varphi_t\left(\tilde{s}_t^i\right)U^i\left(s_{t+1}^i,\lambda\right) \tag{22}$$

where $\varphi_t\left(\tilde{s}_t^i\right)$ is a diminishing step-size. The key idea of the update is that, instead of compute $\tilde{U}\left(\tilde{s}_t^i,\lambda\right)$ as in Eq. (21), we use one realization of $U^i\left(s_{t+1}^i,\lambda\right)$ to represent $\tilde{U}\left(\tilde{s}_t^i,\lambda\right)$ and update $\tilde{U}\left(\tilde{s}_t^i,\lambda\right)$ by averaging all the past realizations. The learning procedure is further illustrated in Figure 5.

### *C. Stochastic subgradient-based resource price update*

From Section V.B, we notice that the subgradient of the dual problem with uniform price is computed as in Eq. (14) which is the expected discounted accumulated resource consumption. Since each wireless user does not know the transition probability, we only use the realized sample path to estimate the subgradient of the dual problem (i.e. using the stochastic subgradient). Specifically, we update the Lagrangian multiplier as follows:

$$\lambda_{k+1} = \left[\lambda_k + \kappa_k\left(\sum_{i=1}^{M}\sum_{t=0}^{\infty}(\alpha)^t\left(x_t^i\right) - \frac{1}{1-\alpha}\right)\right]^+ \tag{23}$$

where $\sum_{t=0}^{\infty}(\alpha)^t\left(x_t^i\right)$ is the stochastic subgradient approximating the subgradient $Z^i$ and $\kappa_k$ is a diminishing step-size. However, in practice, we cannot wait for an infinite time to update the Lagrangian multiplier. Instead, we update the multiplier every $K$ time slots, i.e. we use $\tilde{Z}^i = \sum_{t=kK}^{(k+1)K-1}(\alpha)^{t-kK}\left(x_t^i\right)$ instead of $\sum_{t=0}^{\infty}(\alpha)^t\left(x_t^i\right)$. The proposed online learning algorithm is illustrated in Figure 6.

## VII. SIMULATION RESULTS

In this section, we present simulation results highlighting the efficiency of the proposed single-user and multi-user video transmission solutions compared to existing solutions.

### *A. Single-user video transmission*

To compress the video data, we used a scalable video coding scheme [23], which is attractive for wireless streaming applications because it provides on-the-fly application adaptation to channel conditions, support for a variety of wireless receivers with different resource capabilities and power constraints, and easy prioritization of various coding layers and video packets. In this section, we consider one user transmitting the video sequence “Coastguard” (CIF resolution, 30 Hz) over the time-varying wireless channel, which is modelled as a FSMC with eight channel states (10dB, 15dB, 18dB 20dB, 23dB, 25dB, 28dB and 30dB, respectively). We compare our proposed approach using the foresighted scheduling and resource acquisition policies to a state-of-the-art "distortion impact"-based packet scheduling solution, which only maximizes the current video quality of the frames within the STW, and we refer to this solution as the "conventional priority-based solution" [2][3][4]. Figure 7 shows the average distortion reduction (computed as in Eq. (2)) experienced by the user under various resource constraints (i.e. transmission time allocated to the user). From this figure, we notice that, to receive the same distortion reduction, our proposed solution can save 5%~20% of the resources. The improvement is due to the fact that our solution considers the impact of the environmental dynamics and foresightedly schedules the video data for transmission. We further note that the difference between our proposed solution and conventional priority-based solutions becomes smaller when the network resource is either very scarce or plentiful. This can be explained as follows. On one hand, when the network resource is very scarce, our proposed solution only schedules the most important data (e.g. low-frequency frames in the wavelet-based encoded data), which is the same as the "distortion-impact"-based policy. On the other hand, when the network resource is plentiful, both policies can transmit all the video data. However, in the most usage scenarios, when the resources are neither plentiful nor very scarce (in which case the users may not transmit anyway), Figure 8 shows that our approach improves the video quality by 1.5dB in terms of Peak Signal-to-Noise Ratio (PSNR).

### *B. Dual solutions with uniform price*

In this section, we will verify the convergence of the dual solution with uniform price to the proposed MUMDP. We will further compare the performance of our approach to that of the conventional multi-user dual solution. We first consider three wireless users: User 1 streams the video sequence "Foreman" (CIF resolution, 30 Hz), User 2 streams the video sequence "Coastguard" (CIF resolution, 30 Hz) and User 3 streams the video sequence "Mobile" (CIF resolution, 30 Hz). We compare our proposed dual solution with uniform price to the conventional dual solution [15] based on the NUM framework. Figure 9 shows the convergence of the resource prices with various initial price selections. We notice that, our proposed dual solution with uniform price shows much faster convergence (less than 25 iterations) than the conventional dual solution (having more than 100 iterations). We also note that our solution converges to a lower resource price than the conventional one. This is because that the conventional solution myopically maximizes the video transmission over each time slot. Hence, to achieve a feasible resource allocation, it has to increase its resource price to ensure that the resource allocations over all the states are feasible (corresponding to the worst case scenario.) However, in our solution, we relax the stage resource constraints into the accumulated resource constraint (shown in the problem of MUP/ARC) and the resource price is reduced. However, we scale down the resource acquisition at each multi-user system state to enforce the feasible allocation. Figure 10 shows the distortion reduction of each user with various initial price selections. It demonstrates that our proposed solution gains higher distortion reduction, which is further verified by the received PSNR shown in Figure 11 (i.e. User 1 receives 0.5dB higher PSNR, User 2 receives 1dB higher PSNR and User 3 receives 1.1dB higher PSNR). The improvement is due to the foresighted decisions in our solution, as compared to the myopic decisions in the conventional NUM-based one.

### C. *Online learning*

In this section, we will verify the convergence rate of our proposed online learning algorithm and corresponding impact on the video transmission. We also compare our algorithm to the conventional online learning algorithm [27], which is often used to improve the wireless transmission strategies with unknown dynamics [37]. We consider three wireless users streaming video sequences as in Section **Error! Reference source not found.**. Different from the settings in Subsection **Error! Reference source not found.**, we assume that all the users initially do not have any statistical information about the channel conditions and incoming data, thereby not knowing the state transitions. Using the proposed online learning, the wireless users keep improving their own resource acquisition policy. The resource price is updated every 100 time slots. Figure 12 shows the average distortion reduction received by each user deployed with the proposed online learning and the standard online learning, separately. From this figure, we notice that, compared to the conventional learning algorithm, our proposed method can significant increase the learning curve (i.e. significantly increasing the average received distortion reduction) and dramatically improve the received distortion reduction over time. Figure 13 shows the received video quality (in terms of PSNR) of each user over time when using these two learning algorithms. This figure further confirms that our proposed learning algorithm can improve the video quality of all the users over time. On average, our proposed algorithm improves the video quality of User 1 by 0.9 dB, User 2 by 1.2 dB and User 3 by 1.4dB in terms of PSNR. This improvement is due to the fact that our proposed approach can update the policy at multiple states during one time slot and hence, exhibits a fast convergence rate.

## VIII. CONCLUSIONS

Unlike the conventional formulations for the video transmission over time-varying wireless networks, we systematically formulate the dynamic multi-user video transmission as an MUMDP problem to explicitly consider the heterogeneous video data and dynamic wireless network conditions. This MDP formulation allows the wireless users to make foresighted decisions in order to maximize the long-term utility (i.e. video quality) instead of the immediate reward, which is essential for video applications. The proposed distributed dynamic optimization approach using Lagrangian relaxation with an uniform resource price allows each wireless user to maximize its own video quality given the resource price. To deal with the unknown video characteristics and channel conditions, and to reduce the computation complexity for each user, a novel online reinforcement learning algorithm has been developed, which allows the wireless users to update their transmission policy in multiple states during one time slot, thereby significantly accelerating the learning speed and improving the received video quality.

## Appendix

### A. *Proof of multi-user Bellman's equation using uniform price*

Proof: We prove this by induction.

We define

$$U_0(\boldsymbol{s},\boldsymbol{\lambda}) = \max_{\substack{\boldsymbol{y}^i \in Y^i \\ x^i \geq 0}} \sum_{i=1}^{M} \left[ u_i\left(s^i, \boldsymbol{y}^i, x^i\right) - \lambda x^i + \frac{1}{M}\lambda \right]$$

$$= \sum_{i=1}^{M} \max_{\substack{\boldsymbol{y}^i \in Y^i \\ x^i \geq 0}} \left[ u_i\left(s^i, \boldsymbol{y}^i, x^i\right) - \lambda x^i + \frac{1}{M}\lambda \right] = \sum_{i=1}^{M} U_0^i\left(s^i, \lambda_{\boldsymbol{s}}\right)$$

with

$$U_0^i\left(s^i, \lambda_{\boldsymbol{s}}\right) = \max_{\substack{\boldsymbol{y}^i \in Y^i \\ x^i \geq 0}} \left[ u_i\left(s^i, \boldsymbol{y}^i, x^i\right) - \lambda x^i + \frac{1}{M}\lambda \right].$$

Similarly,

$$U_1(\boldsymbol{s},\boldsymbol{\lambda}) = \max_{\substack{\boldsymbol{y}^i \in Y^i \\ x^i \geq 0 \\ i=1,\cdots,M}} \left\{ \sum_{i=1}^{M} \left[ u_i\left(s^i, \boldsymbol{y}^i, x^i\right) - \lambda x^i + \frac{1}{M}\lambda \right] + \alpha \sum_{\boldsymbol{s}'} \prod_{k=1}^{M} p\left(s^{k\prime} \mid s^k, \boldsymbol{y}^k, x^k\right) \left[ \sum_{i=1}^{M} U_0^i\left(s^{i\prime}, \lambda\right) \right] \right\}$$

$$= \max_{\substack{\boldsymbol{y}^i \in Y^i \\ x^i \geq 0 \\ i=1,\cdots,M}} \left\{ \sum_{i=1}^{M} \left[ u_i\left(s^i, \boldsymbol{y}^i, x^i\right) - \lambda x^i + \frac{1}{M}\lambda + \alpha \sum_{\boldsymbol{s}'} p\left(s^{i\prime} \mid s^i, \boldsymbol{y}^i, x^i\right) U_0^i\left(s^{i\prime}, \lambda\right) \right] \right\}$$

$$= \sum_{i=1}^{M} \max_{\substack{\boldsymbol{y}^i \in Y^i \\ x^i \geq 0}} \left[ u_i\left(s^i, \boldsymbol{y}^i, x^i\right) - \lambda x^i + \frac{1}{M}\lambda + \alpha \sum_{\boldsymbol{s}'} p\left(s^{i\prime} \mid s^i, \boldsymbol{y}^i, x^i\right) U_0^i\left(s^{i\prime}, \lambda\right) \right]$$

$$= \sum_{i=1}^{M} U_1^i\left(s^i, \lambda\right)$$

With

$$U_1^i\left(s^i, \lambda\right) = \max_{\substack{\boldsymbol{y}^i \in Y^i \\ x^i \geq 0}} \left[ u_i\left(s^i, \boldsymbol{y}^i, x^i\right) - \lambda x^i + \frac{1}{M}\lambda + \alpha \sum_{\boldsymbol{s}'} p\left(s^{i\prime} \mid s^i, \boldsymbol{y}^i, x^i\right) U_0^i\left(s^{i\prime}, \lambda\right) \right].$$

Recursively, we have

$$U(\boldsymbol{s},\lambda) = \lim_{n\to\infty} U_n(\boldsymbol{s},\lambda) = \lim_{n\to\infty} \sum_{i=1}^{M} U_n^i\left(s^i, \lambda\right) = \sum_{i=1}^{M} \lim_{n\to\infty} U_n^i\left(s^i, \lambda\right) = \sum_{i=1}^{M} U^i\left(s^i, \lambda\right)$$

Where

$$U^i\left(s^i, \lambda\right) = \max_{y^i, x^i} \left[ u^i\left(s^i, y^i, x^i\right) - \lambda x^i + \frac{1}{M}\lambda + \alpha \sum_{s^{i\prime}} p\left(s^{i\prime} \mid s^i, \boldsymbol{y}^i, x^i\right) U_i\left(s^{i\prime}, \lambda\right) \right]. \blacksquare$$

*B. Proof of subgradient for uniform price*

Proof: For each given $\lambda$, suppose that $x^{i,*}\left(s^i, \lambda\right)$ and $\boldsymbol{y}^{i,*}\left(s^i, \lambda\right)$, $i = 1, \cdots, M$ maximize the dual Bellman's equations in Eq. (13) and hence, maximize the objective in Eq. (10). Then, we have

$$
\begin{aligned}
U^{\lambda',*}(\boldsymbol{s}) &= \sum_{i=1}^{M} \max_{\boldsymbol{y}^i, x^i} \left\{ \left[ u^i\left(s^i, \boldsymbol{y}^i, x^i\right) - \lambda' x^i + \lambda' \frac{1}{M} \right] + \alpha \sum_{\boldsymbol{s}^{i\prime}} p\left(s^{i\prime} \mid s^i, \boldsymbol{y}^i, x^i\right) U_i^{\lambda',*}\left(s^{i\prime}\right) \right\} \\
&\geq \sum_{i=1}^{M} \left\{ \left[ u^i\left(s^i, \boldsymbol{y}^i\left(s^i, \lambda\right), x^i\left(s^i, \lambda\right)\right) - \lambda' x^i + \lambda' \frac{1}{M} \right] + \alpha \sum_{\boldsymbol{s}^{i\prime}} p\left(s^{i\prime} \mid s^i, \boldsymbol{y}^i, x^i\right) U_i^{\lambda',*}\left(s^{i\prime}\right) \right\} \\
&= \sum_{i=1}^{M} \left\{ \begin{array}{r} \left[ u^i\left(s^i, \boldsymbol{y}^i\left(s^i, \lambda\right), x^i\left(s^i, \lambda\right)\right) - \lambda x^i\left(s^i, \lambda\right) + \lambda \frac{1}{M} \right] + \left(\lambda - \lambda'\right)\left(x^i\left(s^i, \lambda\right) - \frac{1}{M}\right) + \\ \alpha \sum_{\boldsymbol{s}^{i\prime}} p\left(s^{i\prime} \mid s^i, \boldsymbol{y}^i, x^i\right) U_i^{\lambda',*}\left(s^{i\prime}\right) \end{array} \right\}
\end{aligned}
$$

Recursively applying this inequality into $U_i^{\lambda',*}\left(s^{i\prime}\right)$, we further have

$$
U^{\boldsymbol{\lambda}',*}(\boldsymbol{s}) \geq
\sum_{i=1}^{M} \left\{ \begin{array}{l}
\left[ u^i\left(s^i, \boldsymbol{y}^i\left(s^i, \lambda\right), x^i\left(s^i, \lambda\right)\right) - \lambda x^i\left(s^i, \lambda\right) + \lambda \frac{1}{M} \right] + \\
\alpha \sum_{\boldsymbol{s}^{i\prime}} p\left(s^{i\prime} \mid s^i, \boldsymbol{y}^i\left(s^i, \lambda\right), x^i\left(s^i, \lambda\right)\right) \left( \left[ u^i\left(s^{i\prime}, \boldsymbol{y}^i\left(s^{i\prime}, \lambda\right), x^i\left(s^{i\prime}, \lambda\right)\right) - \lambda x^i\left(s^{i\prime}, \lambda\right) + \lambda \frac{1}{M} \right] \right) + \\
\left(\lambda - \lambda'\right) \left[ \left(x^i\left(s^i, \lambda\right) - \frac{1}{M}\right) + \alpha \sum_{\boldsymbol{s}^{i\prime}} p\left(s^{i\prime} \mid s^i, \boldsymbol{y}^i, x^i\right) \left(x^i\left(s^{i\prime}, \lambda\right) - \frac{1}{M}\right) \right] + \\
\alpha^2 \sum_{\boldsymbol{s}^{i\prime}} p\left(s^{i\prime} \mid s^i, \boldsymbol{y}^i\left(s^i, \lambda\right), x^i\left(s^i, \lambda\right)\right) \sum_{\boldsymbol{s}^{i\prime\prime}} p\left(s^{i\prime\prime} \mid s^{i\prime}, \boldsymbol{y}^i\left(s^{i\prime}, \lambda\right), x^i\left(s^{i\prime}, \lambda\right)\right) U_i^{\lambda',*}\left(s^{i\prime\prime}\right)
\end{array} \right\}.
$$

Finally, we have

$$
\begin{aligned}
U^{\boldsymbol{\lambda}',*}(\boldsymbol{s}) &\geq U^{\boldsymbol{\lambda},*}(\boldsymbol{s}) + \left(\lambda - \lambda'\right) \sum_{i=1}^{M} \sum_{\boldsymbol{s}^{i\prime}} \sum_{t=0}^{\infty} \alpha^t \left[\left(P^i\right)^t\right]_{s^i \to s^{i\prime}} \left(x^i\left(s^{i\prime}, \lambda\right) - \frac{1}{M}\right) \\
&= U^{\boldsymbol{\lambda},*}(\boldsymbol{s}) + \left(\lambda - \lambda'\right) \left( \sum_{i=1}^{M} \sum_{\boldsymbol{s}^{i\prime}} \sum_{t=0}^{\infty} \alpha^t \left[\left(P^i\right)^t\right]_{s^i \to s^{i\prime}} \left(x^i\left(s^{i\prime}, \lambda\right)\right) - \frac{1}{1-\alpha} \right) \\
&= U^{\boldsymbol{\lambda},*}(\boldsymbol{s}) + \left(\lambda - \lambda'\right) \left( \sum_{i=1}^{M} \boldsymbol{e}_{s^i}^T \left(I - P^i\right)^{-1} \boldsymbol{x}^i(\lambda) - \frac{1}{1-\alpha} \right)
\end{aligned}
$$

where $P^i = \left[p^i\left(s^{i\prime} \mid s^i, \boldsymbol{y}^i\left(s^i, \lambda\right), x^i\left(s^i, \lambda\right)\right)\right]_{s^i \in S^i, s^{i\prime} \in S^i}$, $\left(P^i\right)^0_{s^i \to s^i} = 1$ and $\left(P^i\right)^0_{s^i \to s^{i\prime}} = 0$, $\boldsymbol{e}_{s^i}$ is a vector with the $s^i$ component being 1 and others being zero.

Hence, the subgradient with respect to $\lambda_{s'}$ is given by $\left( \sum_{i=1}^{M} \boldsymbol{e}_{s^i}^T \left(I - P^i\right)^{-1} \boldsymbol{x}^i(\lambda) - \frac{1}{1-\alpha} \right)$. ■

## REFERENCES

[1] M. van der Schaar, and S. Shankar, "Cross-layer wireless multimedia transmission: challenges, principles, and new paradigms," *IEEE Wireless Commun. Mag.*, vol. 12, no. 4, Aug. 2005.

[2] P. Chou, and Z. Miao, "Rate-distortion optimized streaming of packetized media," *IEEE Trans. Multimedia*, vol. 8, no. 2, pp. 390-404, 2005.

[3] M. van der Schaar, and D. Turaga, "Cross-Layer Packetization and Retransmission Strategies for Delay-

Sensitive Wireless Multimedia Transmission," *IEEE Transactions on Multimedia*, vol. 9, no. 1, pp. 185-197, Jan., 2007.

[4] C. D. Vleeschouwer, and P. Frossard, "Dependent packet transmission policies in rate-distortion optimized media scheduling," *IEEE Transactions on Multimedia*, vol. 9, no 6, October 2007, pp. 1241-1258.

[5] Z. Li, F. Zhai, and A.K. Katsaggelos, "Joint Video Summarization and Transmission Adaptation for Energy-Efficient Wireless Video Streaming," *EURASIP Journal on Advances in Signal Processing, special issue on Wireless Video*, vol. 2008.

[6] Y. J. Liang and B. Girod, "Network-Adaptive Low-Latency Video Communication over Best-Effort Networks," *IEEE Transactions on Circuits and Systems for Video Technology,* vol. 16, no. 1, pp. 72-81, January 2006.

[7] R. Hamzaoui, V. Stankovic, and Z. Xiong, "Optimized error protection of scalable image bitstreams," *IEEE Signal Processing Magazine*, vol. 22, pp. 91-107, November 2005.

[8] B. Lamparter, A. Albanese, M. Kalfane, and M. Luby, "PET-priority encoding transmission: a new, robust and efficient video broadcast technology," *Proc. of ACM Multimedia*, 1995.

[9] R. Berry and R. G. Gallager, "Communications over fading channels with delay constraints," *IEEE Trans. Inf. Theory*, vol 48, no. 5, pp. 1135-1149, May 2002.

[10] Q. Liu, S. Zhou, and G. B. Giannakis, "Cross-layer combing of adaptive modulation and coding with truncated ARQ over wireless links," *IEEE Trans. Wireless Commun.*, vol. 4, no. 3, May 2005.

[11] W. Chen, U. Mitra, and M. J. Neely, "Energy-Efficient Scheduling with Individual Packet Delay Constraints over a Fading Channel," *Wireless Networks*, DOI 10.1007/s11276-007-0093-y.

[12] T. Holliday, A. Goldsmith, and P. Glynn, "Optimal Power Control and Source-Channel Coding for Delay Constrained Traffic over Wireless Channels," *Proceedings of IEEE International Conference on Communications*, vol. 2, pp. 831 - 835, May 2002.

[13] M. Chiang, S. H. Low, A. R. Caldbank, and J. C. Doyle, "Layering as optimization decomposition: A mathematical theory of network architectures," *Proceedings of IEEE*, vol. 95, no. 1, 2007.

[14] X. Zhu, P. Agrawal, J. P. Singh, T. Alpcan, and B. Girod, "Rate Allocation for Multi-User Video Streaming over Heterogenous Access Networks," *Proc. ACM Multimedia, MM'07,* Augsburg, Germany, September 2007.

[15] J. Huang, Z. Li, M. Chiang, and A.K. Katsaggelos, "Joint Source Adaptation and Resource Allocation for Multi-User Wireless Video Streaming," *IEEE Trans. Circuits and Systems for Video Technology*, vol. 18, issue 5, 582-595, May 2008.

[16] E. Maani, P. Pahalawatta, R. Berry, T.N. Pappas, and A.K. Katsaggelos, "Resource Allocation for Downlink Multiuser Video Transmission over Wireless Lossy Networks," *IEEE Transactions on Image Processing*, vol. 17, issue 9, 1663-1671, September 2008.

[17] G-M. Su, Z. Han, M. Wu, and K.J.R. Liu, "Joint Uplink and Downlink Optimization for Real-Time Multiuser Video Streaming Over WLANs," *IEEE Journal of Selected Topics in Signal Processing*, vol. 1, no. 2, pp. 280-294, August 2007.

[18] F. Fu and M. van der Schaar, "Noncollaborative Resource Management for Wireless Multimedia Applications Using Mechanism Design," *IEEE Trans. Multimedia*, vol. 9, no. 4, pp. 851-868, Jun. 2007.

[19] D. P. Bertsekas, "Dynamic programming and optimal control," 3rd, Athena Scientific, Belmont, Massachusetts, 2005.

[20] J. Hawkins, "A Lagrangian decomposition approach to weakly coupled dynamic optimization problems and its applications," PhD Dissertation, MIT, Cambridge, MA, 2003.

[21] D. Adelman, and A. J. Mersereau, "Relaxation of weakly coupled stochastic dynamic programs," *Operations Research*, vol. 56, no. 3, pp. 712-727, May-June 2008.

[22] T. Wiegand, G. J. Sullivan, G. Bjontegaard, A. Luthra, "Overview of the H.264/AVC video coding standard," *IEEE Transactions on Circuits and Systems for Video Technology*, vol. 13, no. 7, pp. 560-576, July, 2003.

[23] J.R. Ohm, "Three-dimensional subband coding with motion compensation", *IEEE Trans. Image Processing*, vol. 3, no. 5, Sept 1994.

[24] Q. Zhang, S. A. Kassam, "Finite-state Markov Model for Reyleigh fading channels," *IEEE Trans. Commun.* vol. 47, no. 11, Nov. 1999.

[25] "IEEE 802.11e/D5.0, wireless medium access control (MAC) and physical layer (PHY) specifications: Medium access control (MAC) enhancements for Quality of Service (QoS), draft supplement," June 2003.

[26] D. P. Bertsekas, "Nonlinear programming," Belmont, MA: Athena Scientific, 2nd Edition, 1999.

[27] R. S. Sutton, and A. G. Barto, "Reinforcement learning: an introduction," Cambridge, MA:MIT press, 1998.

[28] D. S. Turaga and T. Chen, "Hierarchical Modeling of Variable Bit Rate Video Sources," *Packet Video*, 2001.

[29] Q. Li, Y. Andreopoulos, and M. van der Schaar, "Streaming-Viability Analysis and Packet Scheduling for Video over QoS-enabled Networks," *IEEE Trans. Veh. Technol*., vol. 56, no. 6, pp. 3533-3549, Nov. 2007.

[30] D. Djonin and V. Krishnamurthy, "Transmission Control in Fading Channels -- A Constrained Markov Decision Process Formulation with Monotone Randomized Policies," *IEEE Transactions Signal Processing*, Vol.55, No.10, pp. 5069--5083, October 2007.

[31] D. Djonin and V. Krishnamurthy, "Q-Learning Algorithms for Constrained Markov Decision Processes with Randomized Monotone Policies: Applications to MIMO Transmission Control," *IEEE Transactions Signal Processing*, Vol.55, No.5, pp.2170--2181, 2007.

[32] D. M. Topkis, "Supermodularity and complementarity," Princeton University Press, Princeton, NJ, 1998.

[33] A. Reibman, and A. Berger, "Traffic descriptors for VBR video teleconferencing over ATM networks," *IEEE/ACM Transactions on Networking*, vol. 3, no. 3, pp. 329-339, 1995.

[34] V. Borkar, and V. Konda, "The actor-critic algorithm as multi-time-scale stochastic approximation," *Sadhana*, vol. 22, part 4, pp. 525-543, Aug. 1997.

[35] F. Fu and M. van der Schaar, "A New Systematic Framework for Autonomous Cross-Layer Optimization," *IEEE Trans. Veh. Tech.*, to appear.

[36] D. Bertsekas, and R. Gallager, "Data networks," Prentice Hall, Inc. Upper Saddle River, NJ, 1987.

[37] C. Pandana, and K. J. R. Liu, "Near-optimal reinforcement learning framework for energy-aware sensor communications," *IEEE J. Select. Areas Commun.* vol. 23, no. 4, April, 2005.

Figures and Tables

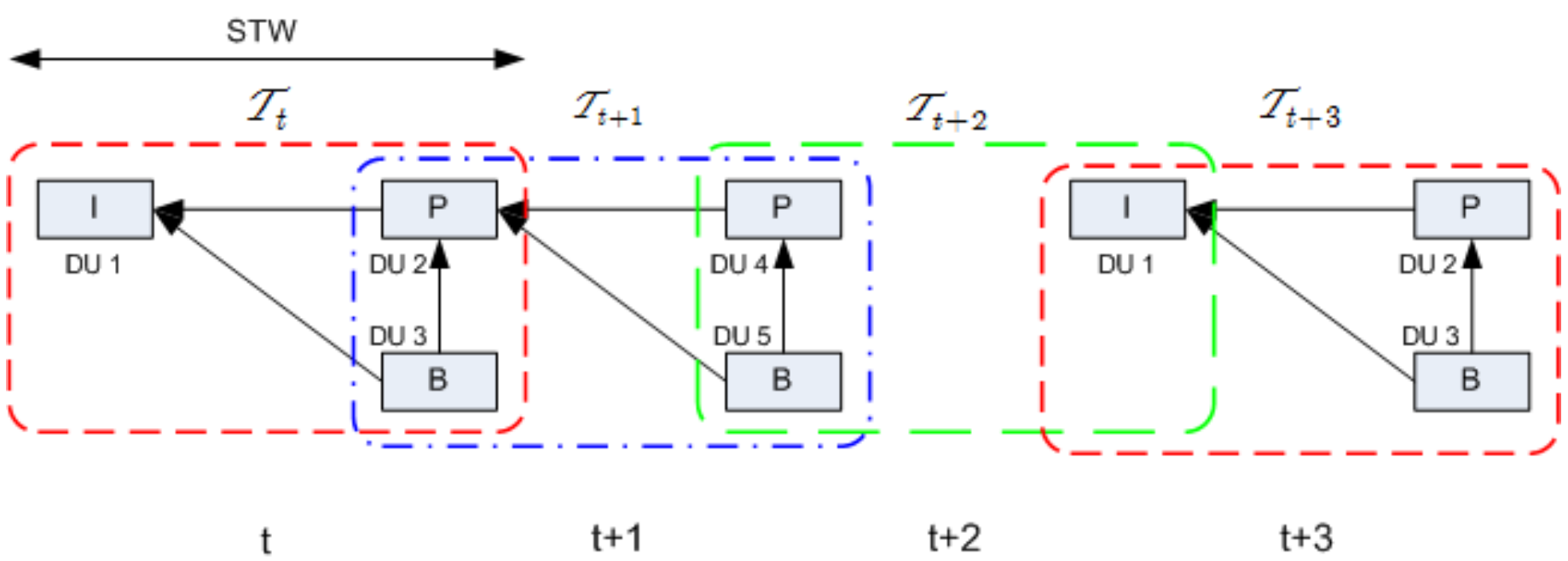

Figure 1. DAG-based dependencies and traffic states at each time slot using IBPBP GOP structure

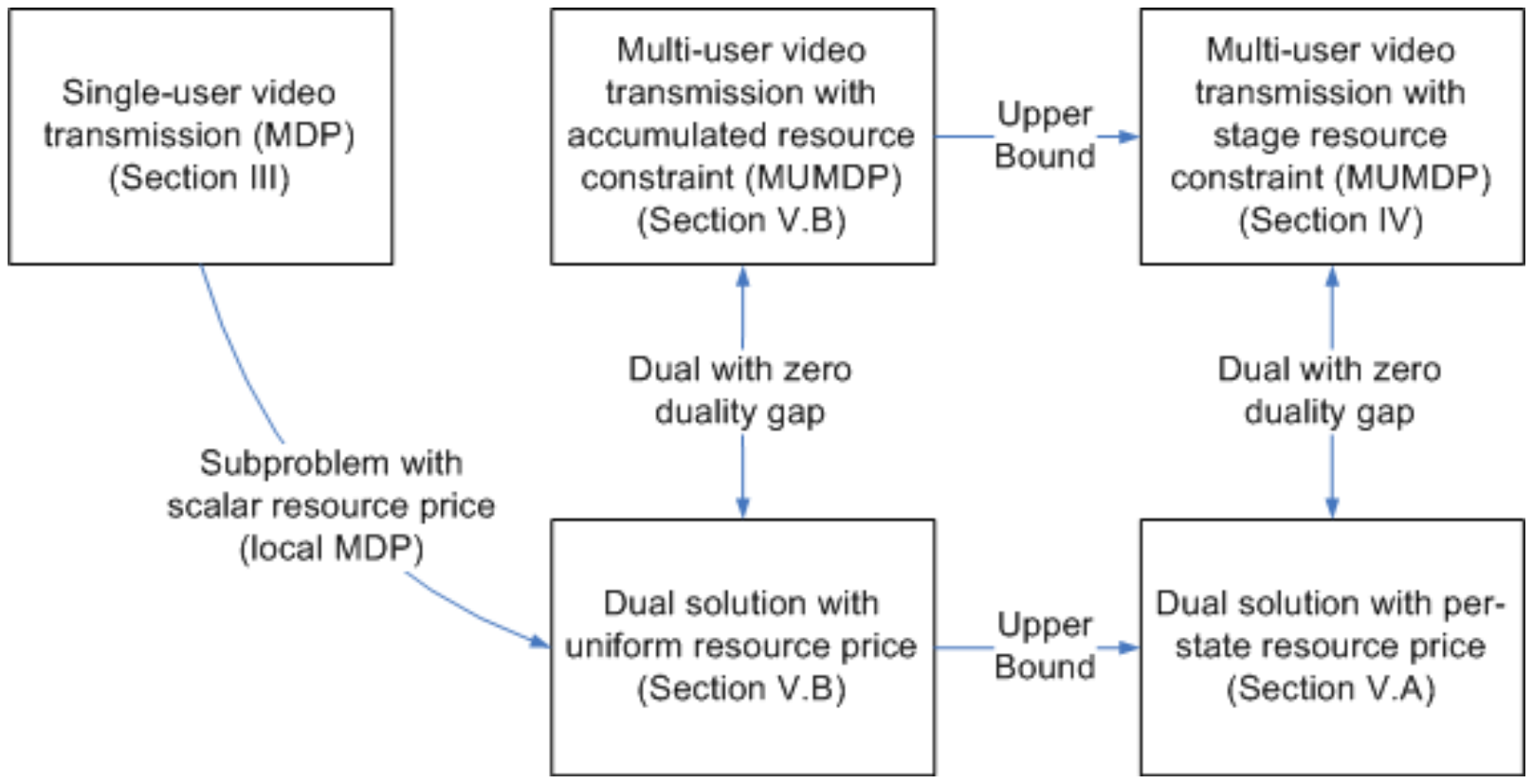

Figure 2. Relationship between the various proposed solutions for the considered multi-user MDP problem

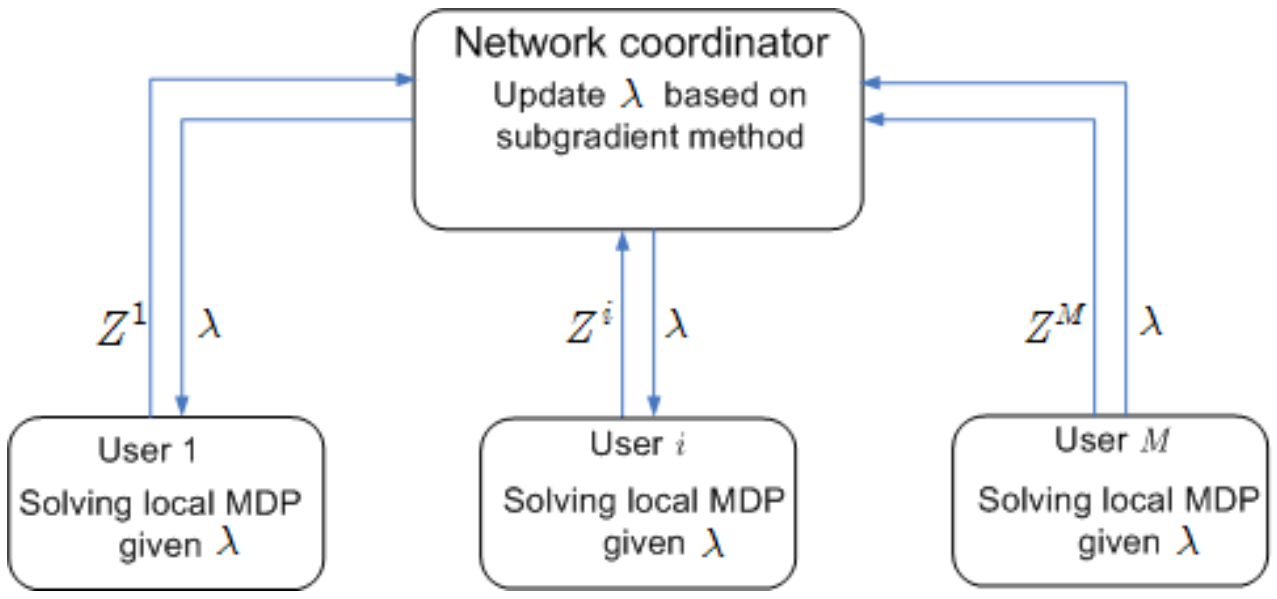

Figure 3. Message exchange in the dual method with uniform price for the considered multi-user wireless video transmission

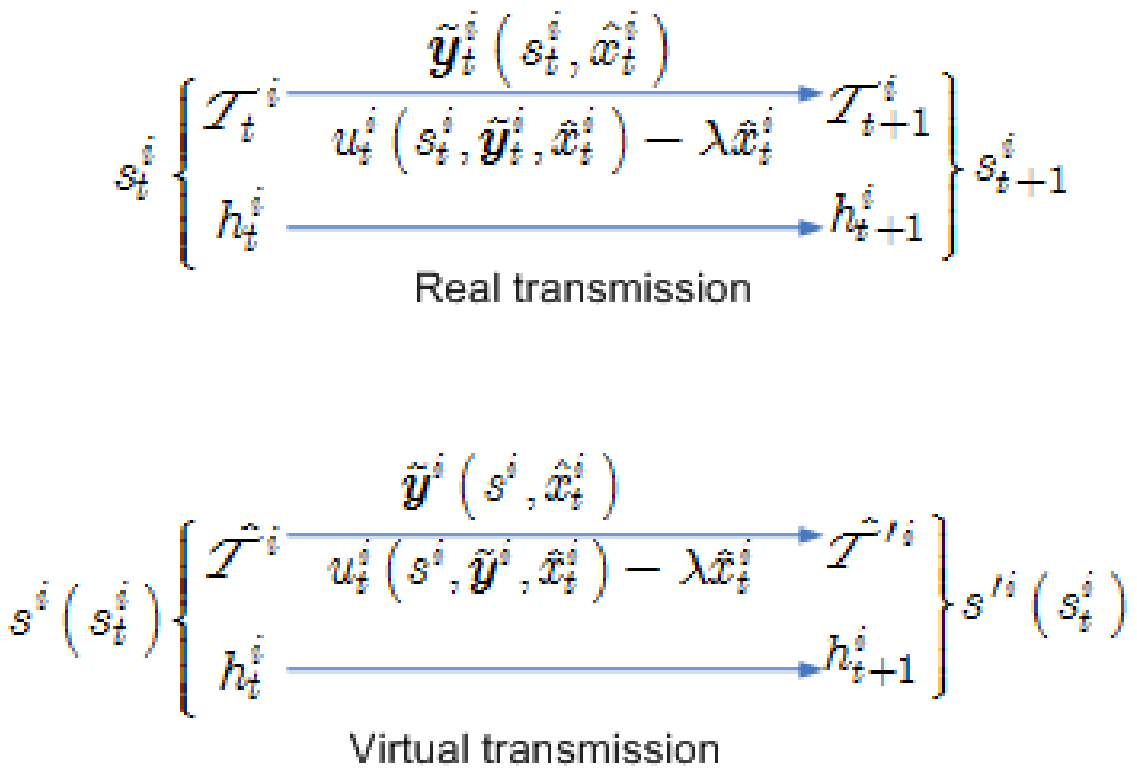

Figure 4. Associated state transition in the online learning given the scheduling and resource allocation

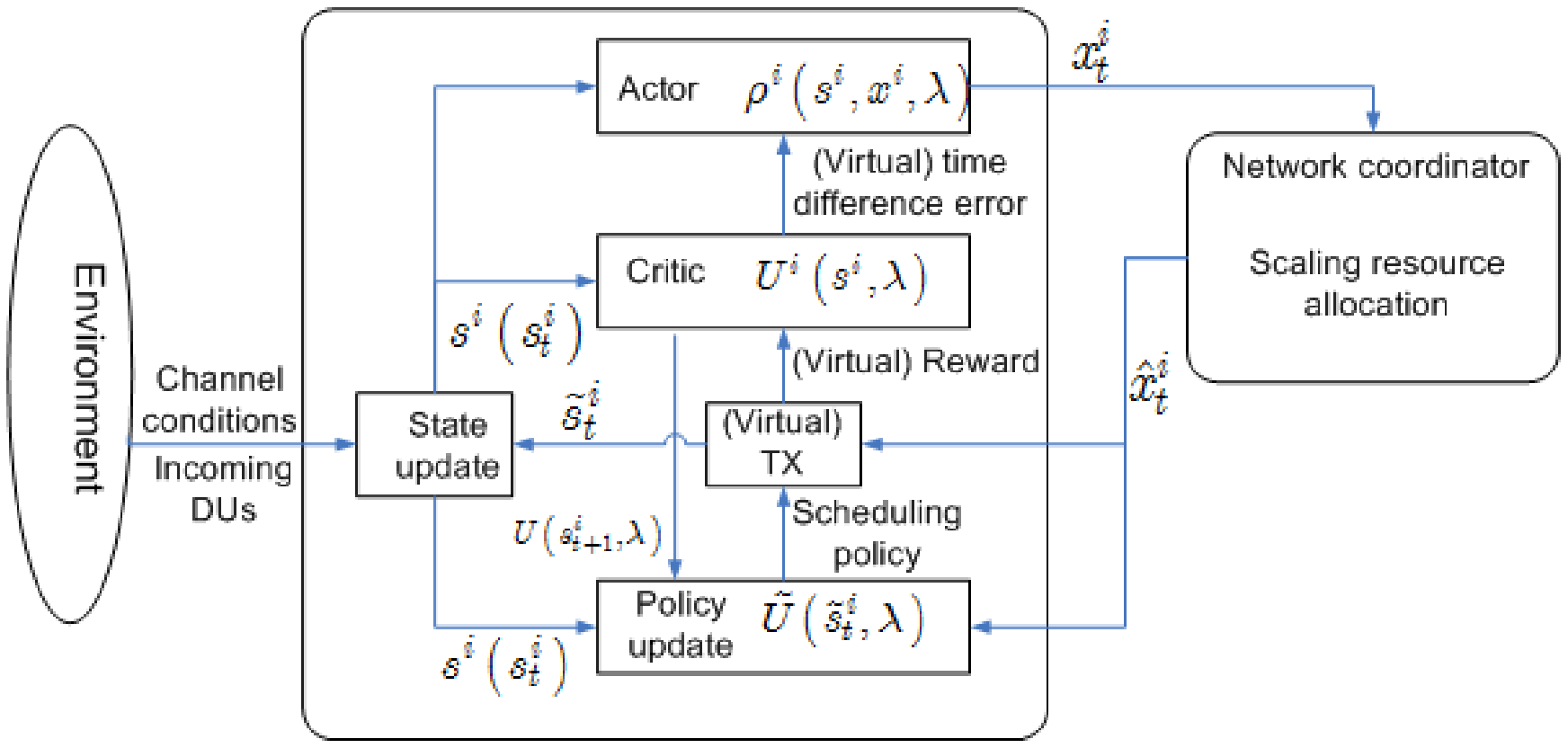

Figure 5. Procedures of the modified Actor-Critic learning within one user

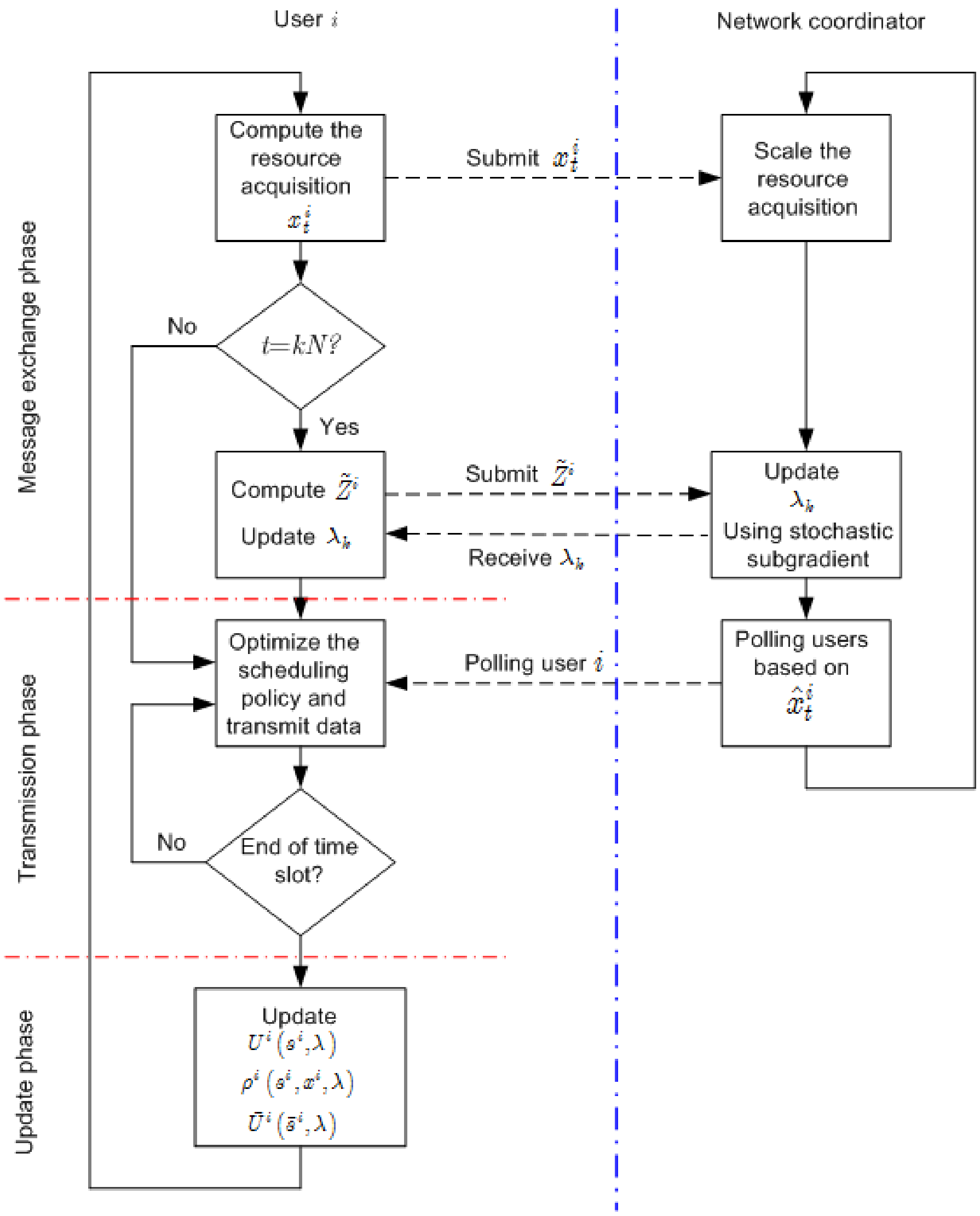

Figure 6. Flowchart of the online learning algorithm

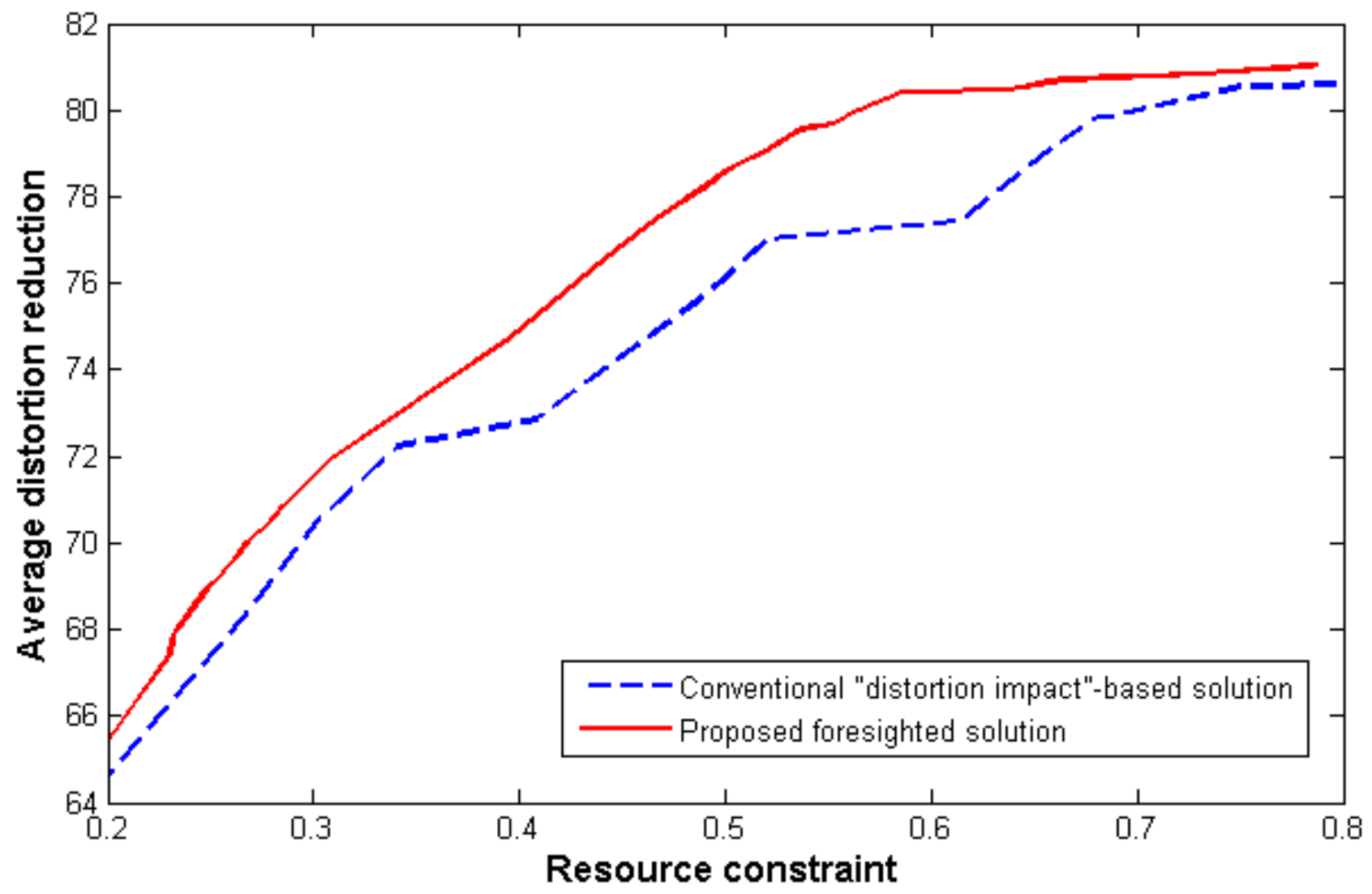

Figure 7. Distortion reduction vs. consumed resource using various approaches

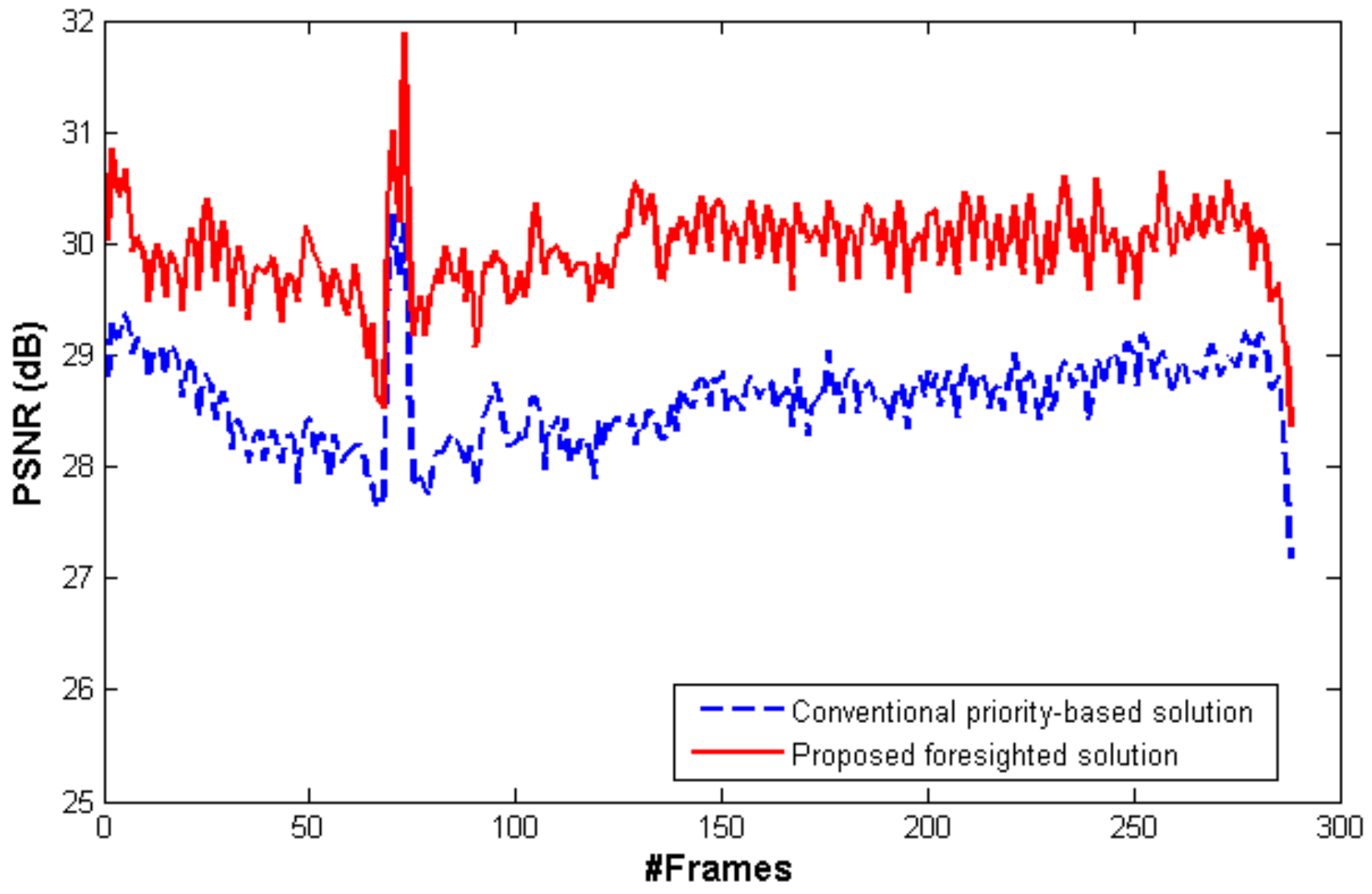

Figure 8. Received PSNR using various approaches

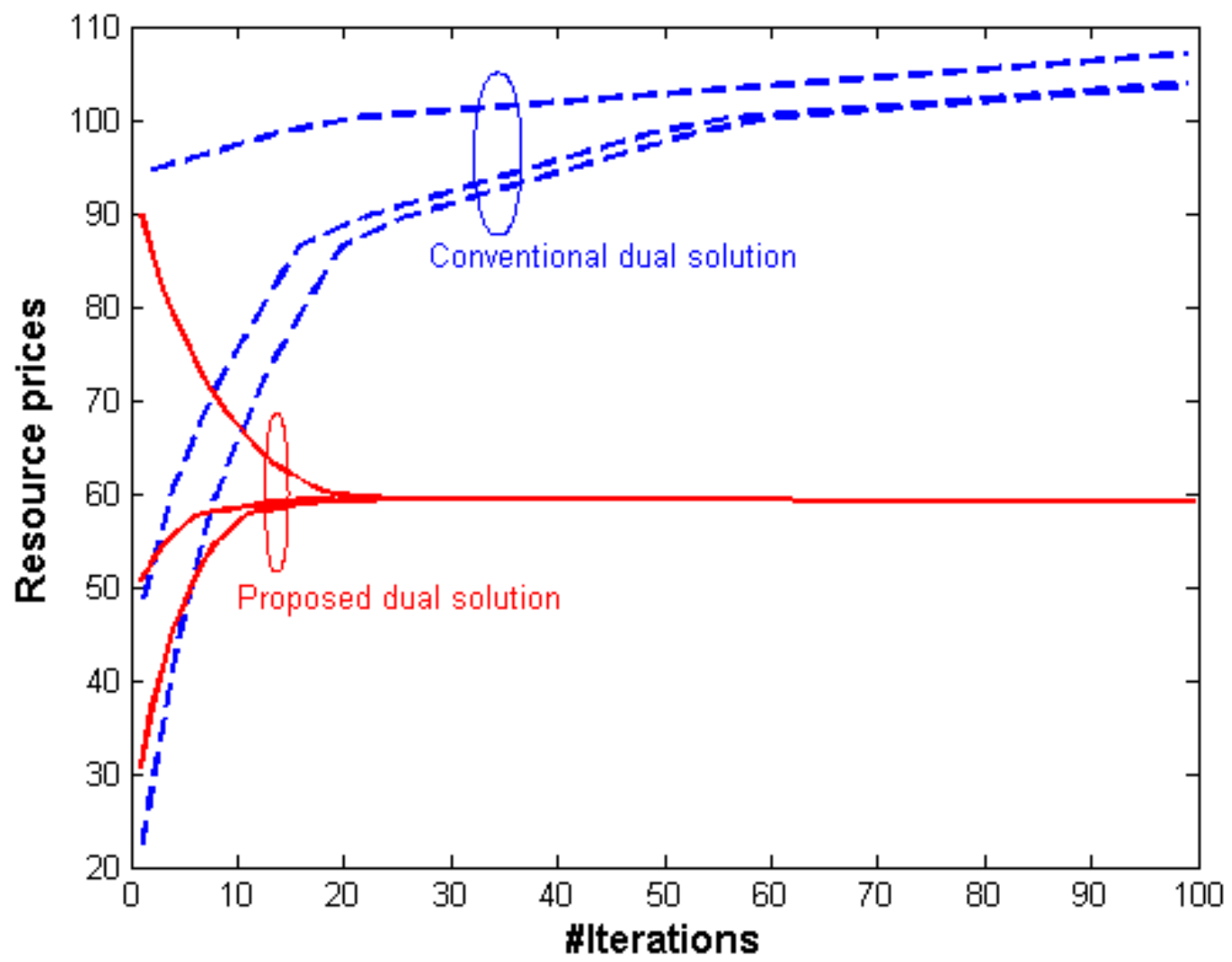

Figure 9. Convergence of the dual solutions under various initial resource price selection

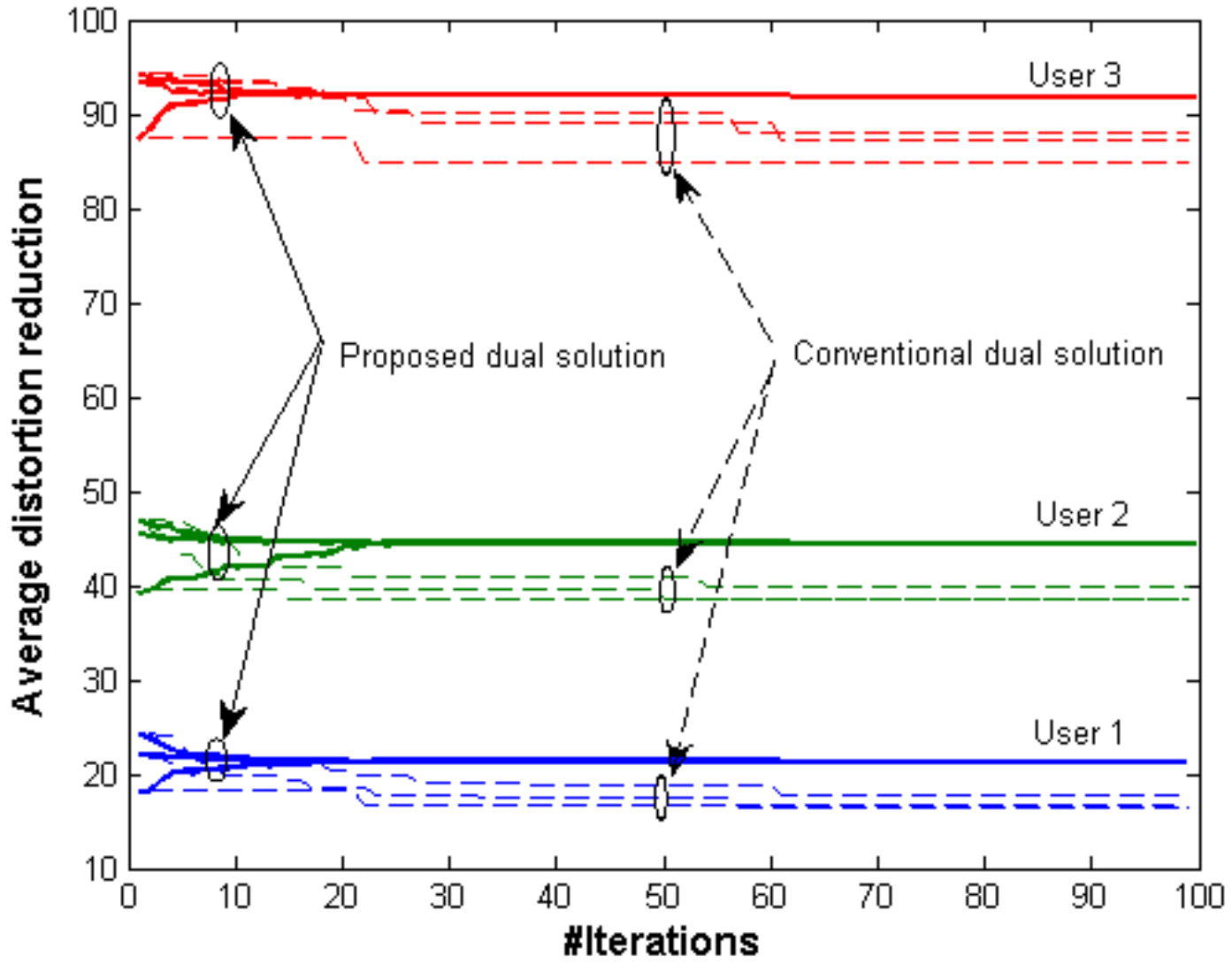

Figure 10. Distortion reduction of each user with the dual solutions under various initial resource price selection

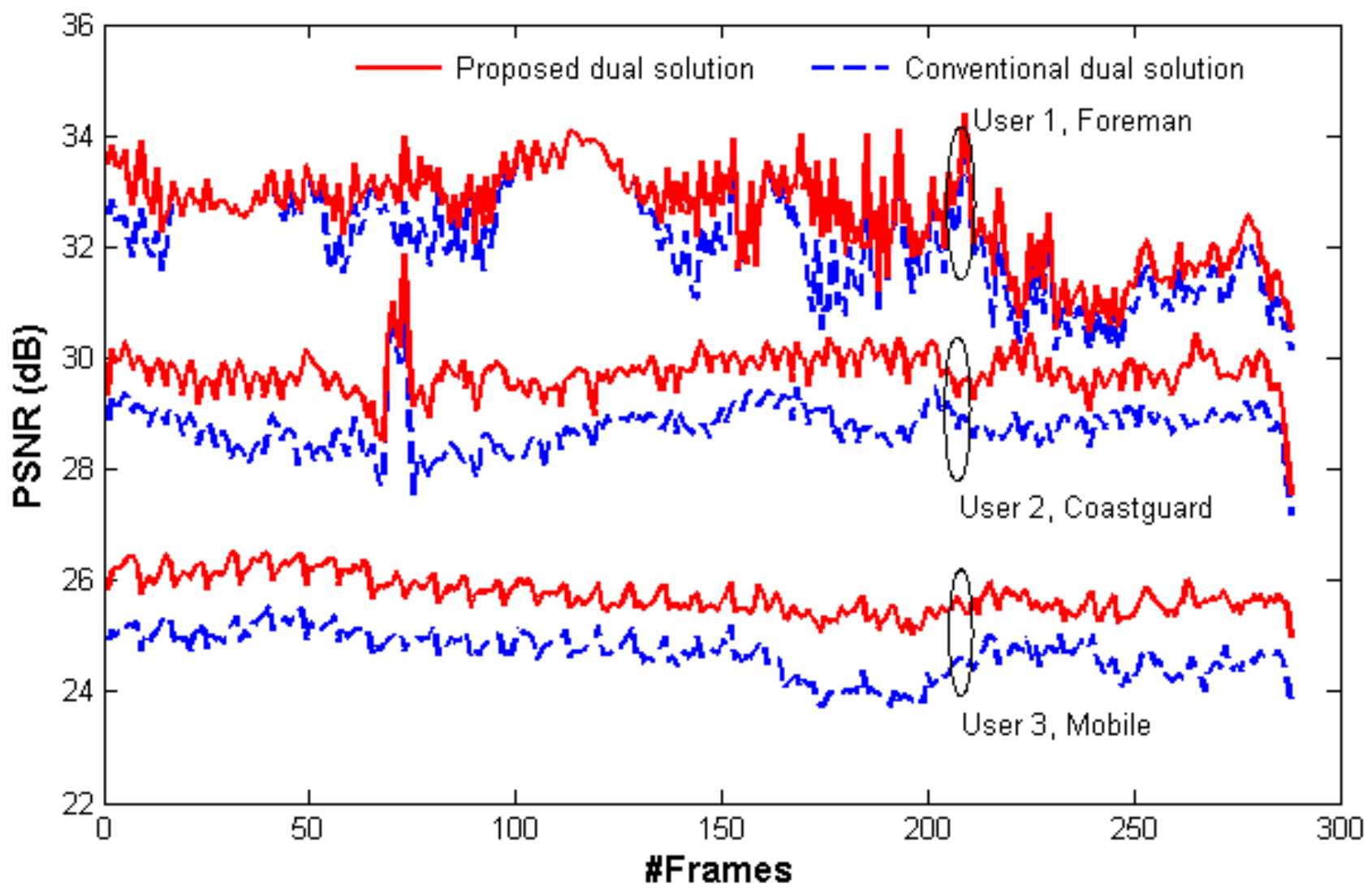

Figure 11. PSNR of streamed video sequences by each user with various solutions

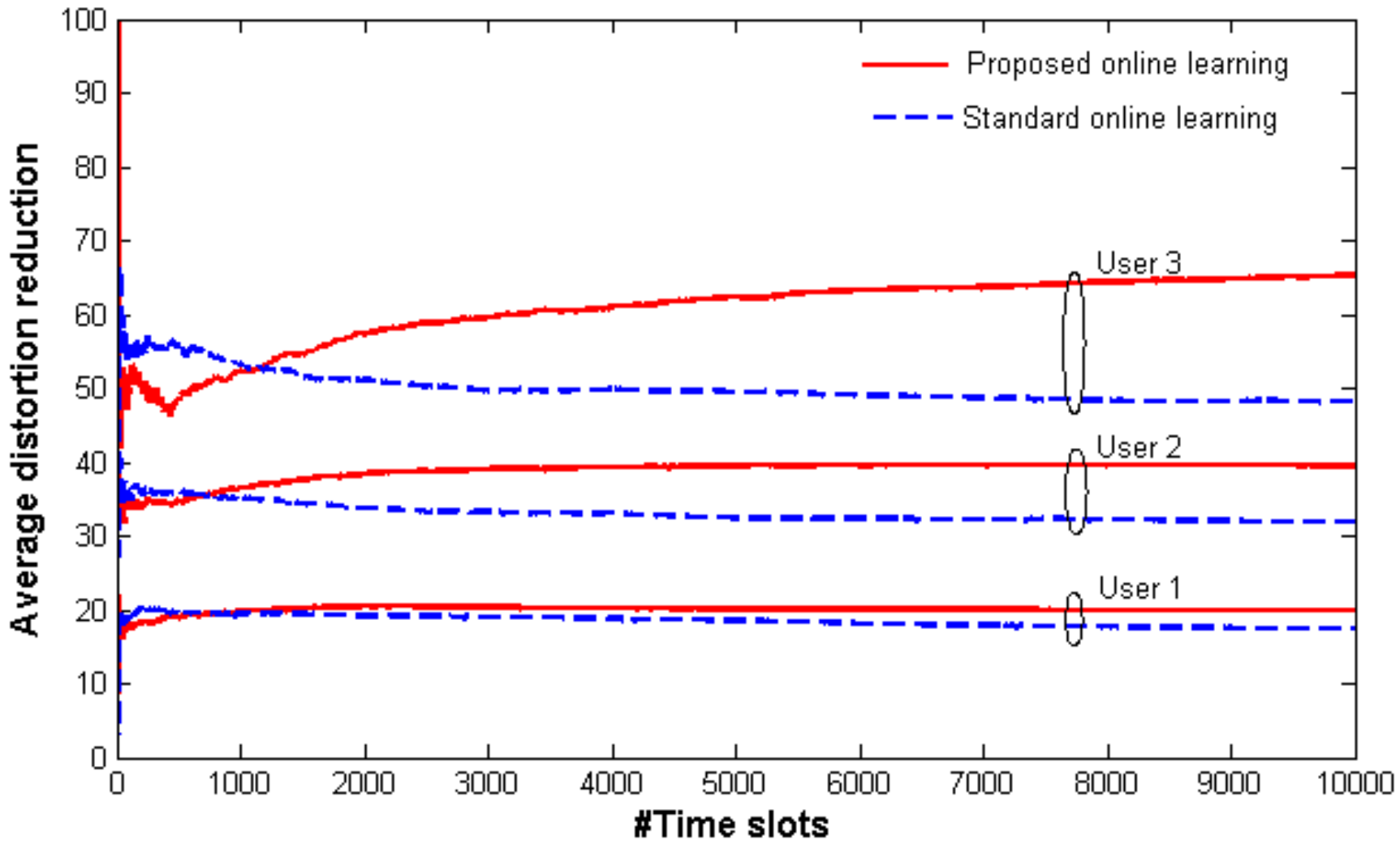

Figure 12. Learning curves of each user with different online learning algorithms

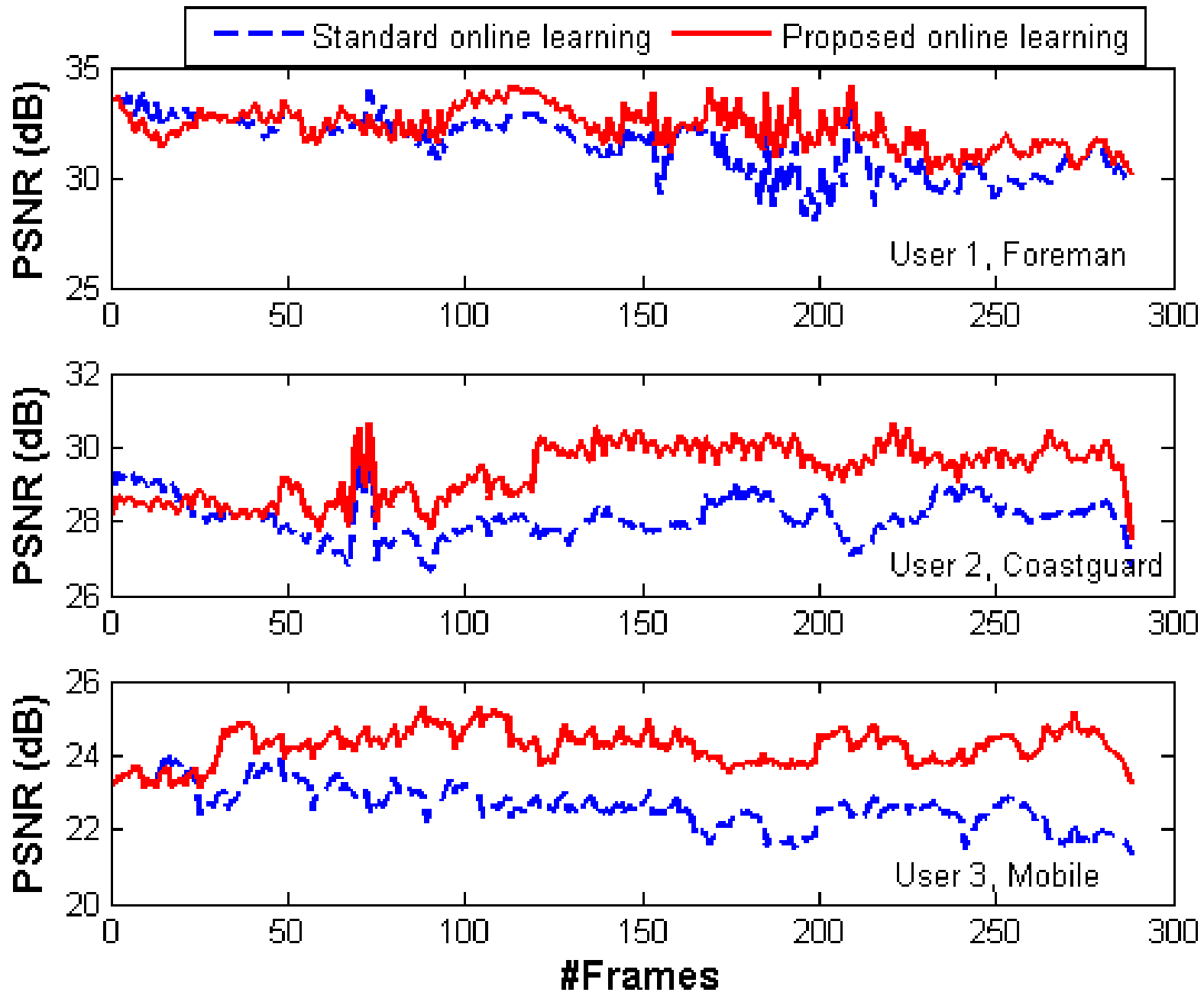

Figure 13. Received PSNR of each user with different online algorithms